\documentclass[aps,prd,showpacs,floatfix,nofootinbib,twocolumn,10pt]{revtex4}

\usepackage{color,amsmath,amssymb,graphicx,latexsym,subfigure}
\usepackage{epstopdf,threeparttable,units,txfonts,ulem,comment}

\newcommand{\sv}{\ensuremath{\langle\sigma v\rangle}}

\begin{document}

\title{Dark Matter Particle Explorer observations of high-energy 
cosmic ray electrons plus positrons and their physical implications}

\author{Qiang Yuan$^{a,b,c}$\footnote{yuanq@pmo.ac.cn}}
\author{Lei Feng$^{a}$}

\affiliation{
$^a$Key Laboratory of Dark Matter and Space Astronomy, Purple Mountain 
Observatory, Chinese Academy of Sciences, Nanjing 210008, China\\
$^a$School of Astronomy and Space Science, University of Science and 
Technology of China, Hefei 230026, China\\
$^c$Center for High Energy Physics, Peking University, Beijing 100871, China
}


\begin{abstract}

The DArk Matter Particle Explorer (DAMPE) is a satellite-borne, 
high-energy particle and $\gamma$-ray detector, which is dedicated to 
indirectly detecting particle dark matter and studying high-energy 
astrophysics. The first results about precise measurement of the
cosmic ray electron plus positron spectrum between 25 GeV and 4.6
TeV were published recently. The DAMPE spectrum reveals an interesting 
spectral softening arount $0.9$ TeV and a tentative peak around $1.4$ 
TeV. These results have inspired extensive discussion. 
The detector of DAMPE, the data analysis, and the first results are
introduced. In particular, the physical interpretations of the DAMPE 
data are reviewed.

\end{abstract}

\pacs{95.35.+d,95.85.Ry,96.50.S-,96.50.sb}

\maketitle

\section{Introduction}

The existence of dark matter (DM) in our Universe is well established 
by many astronomical and cosmological observations (see reviews
\cite{1996PhR...267..195J,2005PhR...405..279B}). The physical nature
of DM, which is largely unknown, becomes one of the most important
and fundamental questions of modern physics. The evolution of large 
scale structures of the Universe and the relic abundance of DM
indicates that DM is most probably a kind of weakly interacting
massive particle (WIMP) beyond the standard model of particle 
physics. This postulated weak interaction feature of DM makes it
detectable by particle detectors. In general there are three ways
proposed to detect WIMP DM: the direct detection of the WIMP-nucleus
scattering by underground detectors, the indirect detection of the
annihilation/decay of WIMPs, and the production of WIMP pairs in
large particle colliders \cite{2005PhR...405..279B,2013FrPhy...8..794B}. 
Quite a number of experiments have been carried out to search for 
WIMPs since the 1980s \cite{1990PhR...187..203S,2013FrPhy...8..794B}.

While the direct detection experiments keep on pushing the WIMP-nucleon
scattering cross section lower and lower \cite{2017PhRvL.118b1303A,
2017PhRvL.119r1301A,2017PhRvL.119r1302C}, interesting anomalies 
have been found in cosmic ray (CR) and $\gamma$-ray observations
in recent years \cite{2008Natur.456..362C,2009Natur.458..607A,
2009PhRvL.102r1101A,2012PhRvL.108a1103A,2013PhRvL.110n1102A,
2014PhRvL.113v1102A,2011PhLB..697..412H,2017ApJ...840...43A}. 
Among these anomalies, the biggest puzzle is probably the positron 
excess discovered\footnote{See also earlier hints from HEAT and AMS 
measurements \cite{1997ApJ...482L.191B,2007PhLB..646..145A}.} by 
PAMELA \cite{2009Natur.458..607A}, Fermi-LAT \cite{2012PhRvL.108a1103A}, 
and AMS-02 \cite{2013PhRvL.110n1102A}, and the associated total 
cosmic ray electron plus positron (CRE) excess by ATIC 
\cite{2008Natur.456..362C}, Fermi-LAT \cite{2009PhRvL.102r1101A},
and AMS-02 \cite{2014PhRvL.113v1102A}. The positron and CRE excesses
suggest the existence of {\tt primary} positron and electron sources
besides the {\tt secondary} component from inelastic collisions of
CR nuclei and the interstellar medium. No significant excess of
CR antiprotons \cite{2009PhRvL.102e1101A,2010PhRvL.105l1101A,
2016PhRvL.117i1103A} constrains that the {\tt primary} sources of these 
CREs should be ``leptonic'' (see, e.g., \cite{2010IJMPD..19.2011F,
2012APh....39....2S,2013FrPhy...8..794B}). Two leading scenarios to 
explain the positron and CRE excesses are pulsars 
\cite{1970ApJ...162L.181S,2009JCAP...01..025H,2009PhRvL.103e1101Y}
and DM annihilation/decay \cite{2008PhRvD..78j3520B,2009NuPhB.813....1C,
2009PhRvD..79b3512Y}. The DM scenario is severely constrained by 
the $\gamma$-ray and radio observations \cite{2009JCAP...03..009B,
2009PhRvD..79h1303B,2010JCAP...03..014P,2010NuPhB.840..284C}, 
leaving pulsars the most likely origin of these excesses. However,
the recent observations of TeV $\gamma$-ray emission from two nearby
pulsars by HAWC indicated that the CREs from these two pulsars were 
not enough to account for the positron excess \cite{2017Sci...358..911A}, 
unless a special transportation of CREs was assumed 
\cite{2017PhRvD..96j3013H,2018arXiv180302640F}. The origin of the 
positron and CRE excesses is still not clear yet.

More precise measurements of the spectral behaviors and anisotropies 
of positrons and/or CREs are particularly helpful in identifying their 
origin. For example, the fine structures of the CRE spectrum may be 
used to distinguish pulsar and DM models \cite{2009PhLB..681..220H,
2009PhRvD..80f3005M,2010JCAP...12..020P,2013PhRvD..88b3001Y}.
The DArk Matter Particle Explorer (DAMPE) is dedicated to precise
observations of CREs, $\gamma$-rays, and CR nuclei in space
\cite{ChangJin:550,2017APh....95....6C}. The major science of
DAMPE is to explore the TeV window via CREs and $\gamma$-rays,
and to search for possible DM signals. The $\gamma$-ray and nucleus
data of DAMPE also enable us to study high-energy astrophysical
phenomena and the CR physics \cite{2017APh....95....6C}.

The DAMPE mission was launched into a 500 km Sun-synchronous orbit on 
December 17, 2015. It has operated smoothly on orbit for more than two
years, and has recorded over 4 billion CR events, with a daily event
rate of $\sim5$ million. All detectors of DAMPE operate perfectly as
expected, which brings us new views about the high-energy Universe.
The first results about the CRE spectrum based on the first 530 days 
of data have been published recently \cite{2017Natur.552...63D}. 
This review introduces the DAMPE mission, the first results, and the 
proposed physical interpretations (see also Ref.~\cite{2018arXiv180606534G}
for a mini-review).

\section{DAMPE experiment}

\begin{figure}[!htb]
\centering
\includegraphics[width=0.48\textwidth]{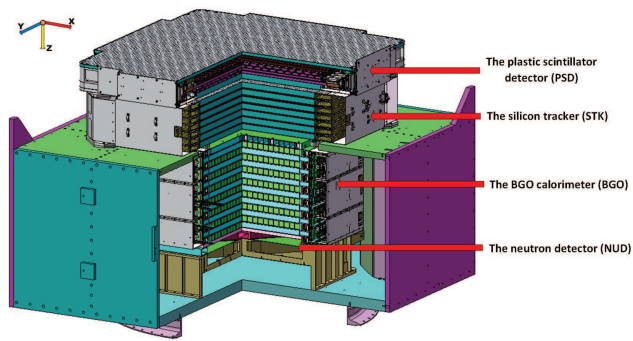}
\caption{Layout of the DAMPE detector. Plot from 
Ref.~\cite{2017APh....95....6C}.
\label{fig:detector}}
\end{figure}

DAMPE is a calorimetric-type, satellite-borne detector. The DAMPE instrument
consists of 4 sub-detectors: from top to bottom, the Plastic Scintillator 
strip Detector (PSD; \cite{2017APh....94....1Y}), the Silicon-Tungsten 
tracKer-converter (STK; \cite{2016NIMPA.831..378A}), the BGO imaging 
calorimeter \cite{2015NIMPA.780...21Z}, and the NeUtron Detector (NUD; 
\cite{2016AcASn..57....1H}). The layout of the detector is shown in 
Fig.~\ref{fig:detector} \cite{2017APh....95....6C}.

The PSD measures the (absolute) charge of an incident particle via the 
ionization effect, and serves as an anti-coincidence detector for 
$\gamma$-rays. The PSD consists of two layers of orthogonally placed 
plastic scintillator bars, with an effective area of $82.5\times82.5$ 
cm$^2$ \cite{2017APh....95....6C}. In each layer, 41 bars are arranged in 
two sub-layers with a shift of 0.8 cm to enable a full coverage of incident 
particles. The overall efficiency of the PSD is about 0.99999 based on 
Monte Carlo (MC) simulations \cite{2018RAA....18...27X}. The charge 
resolution of protons, Carbon, and Iron nuclei is found to be about 
$0.14$ (full width at half maximum), $0.18$ (Gaussian width), and 
$0.32$ (Gaussian width), respectively \cite{Dong-RAA-2018}.

The STK measures the trajectory and the charge of a particle. The STK
consists of 6 double-layers of silicon tracker, each layer with two 
sub-layers arranged orthogonally to measure the $x$ and $y$ positions
of a track. Each silicon layer is made of 16 ladders, each formed by
4 silicon micro-strip detectors (SSD). Each ladder is segmented into
768 strips, half of which are readout strips to limit the number of
readout channels. In total there are $384\times16\times12=73728$
readout channels of the STK. In addition, three 1 mm thick tungsten 
plates are inserted to increase the photon conversion rate. The total 
thickness of the STK is about one radiation length. The spatial
resolution of the STK is better than $50~\mu$m after the alignment
procedure \cite{2017arXiv171202739T}. The angular resolution of
the STK based on MC simulations is about $1^{\circ}.2$ ($0^{\circ}.1$) 
at 1 (100) GeV for normal incident photons \cite{2017APh....95....6C}.
The STK can also measure the particle charge through measuring the 
ionization energy loss. The charge resolution is found to be about 0.04 
for protons and 0.07 for Helium nuclei (both are Gaussian widths;
\cite{DAMPE-STK-charge}). The STK is helpful in improving the charge 
resolution of light nuclei. However, it would get saturated when $Z \gtrsim 8$.

The BGO calorimeter is the major detector of DAMPE instrument which
measures the energy and trajectory of an incident particle, and most 
importantly, provides high-performance electron/hadron discrimination 
based on the shower images. The BGO calorimeter is made of 308 BGO
crystals, placed orthogonally in 14 layers. Each crystal is read out 
at the two ends by photomultiplier tubes coupled to optical filters 
with different attenuation factors. At each end the signal is read 
out from three dynodes with different gain. Such a design can enlarge 
the dynamic range and check the consistency of the energy measurement
\cite{2015NIMPA.780...21Z}. The effective area of the calorimeter is 
$60\times60$ cm$^2$, and the thickness is about 32 radiation lengths. 
This very thick calorimeter enables full development of electro-magnetic
showers with primary energies up to 10 TeV without significance leakage
\cite{2017NIMPA.856...11Y}. A small fraction of energy leakage ($\sim7\%$) 
due to dead materials can be corrected with a shower-development-dependent 
method \cite{2017NIMPA.856...11Y}. The MC simulations show that the
energy resolution of electrons and photons can be parameterized as
$\sigma/E=\sqrt{0.0073^2 + 0.0618^2/(E/{\rm GeV})}$, which is verified
by the beam test data from 0.5 GeV to 243 GeV \cite{2016NIMPA.836...98Z}.
The energy resolution is about $6\%$ at 1 GeV and $1\%$ above 100 GeV.
The absolute energy scale is determined by the geomagnetic cutoff of the 
CRE spectrum, which gives $1.25\%\pm1.75\%~({\rm stat.})\pm1.34\%~
({\rm sys.})$ higher of the energy scale \cite{Zang-ICRC2017-197}. 
The linearity of the energy measurement has been well confirmed by the 
beam test data up to 243 GeV \cite{2017APh....95....6C}.
For protons the energy resolution is about $10\%\sim30\%$ for energies 
from 10 GeV to 100 TeV \cite{2017APh....95....6C}. The proton rejection 
fraction is found to be higher than $99.99\%$ when keeping $90\%$ of 
CREs based on a two-parameter representation of shower images 
\cite{2017Natur.552...63D}.

At the bottom is the NUD detector which provides additional electron/hadron 
discrimination, due to the fact that a hadronic shower typically gives
one order of magnitude higher number of neutrons than that of an 
electromagnetic shower. The NUD is made of four $30~{\rm cm}\times 
30~{\rm cm}\times 1~{\rm cm}$ plastic scintillators with $5\%$ (weight) 
Boron element. The active area of the NUD is about $61~{\rm cm}\times
61~{\rm cm}$. Neutrons are captured via the reaction $^{10}{\rm B}+n\to
^7{\rm Li}+\alpha$, and the ionization energy loss due to charged Lithium
and $\alpha$ particles can be recorded by photomultiplier tubes. 
Simulations show that the proton rejection power for NUD can reach a 
factor of $\sim10$ for TeV energies \cite{2017APh....95....6C}.

These four sub-detectors enable good measurements of the charge ($|Z|$), 
arrival direction, energy, and particle identity of each event. More
detailed description of the detector and its simulated performance as
well as beam test results can be found in Ref. \cite{2017APh....95....6C}. 

\section{Observations of CREs}

The observations of CREs are difficult, since there are large amount of
proton and nucleus backgrounds. At GeV energies, the flux of protons is
about 100 times higher than that of electrons. The ratio increases with
energies, reaching $\sim1000$ at TeV energies. Any reliable detection
of CREs must effectively reject the proton (or more generally, nucleus)
background, to a level of $10^4$ or more. Usually the magnetic spectrometer
is used for the identification of CREs, such as balloon-borne experiments
HEAT, CAPRICE, and space experiments PAMELA, AMS. Another way is to use 
the imaging calorimeter to distinguish electrons from hadrons, e.g.,
balloon-borne experiments BETS, ATIC, and space experiments Fermi, 
DAMPE, CALET. The ground-based imaging atmospheric Cherenkov telescope 
(IACT) arrays or air shower detector arrays with muon detectors can 
also distinguish CREs from the hadronic background and measure the CRE
spectrum, although there are relatively large systematic uncertainties.

\begin{figure*}[!htb]
\centering
\includegraphics[width=0.48\textwidth]{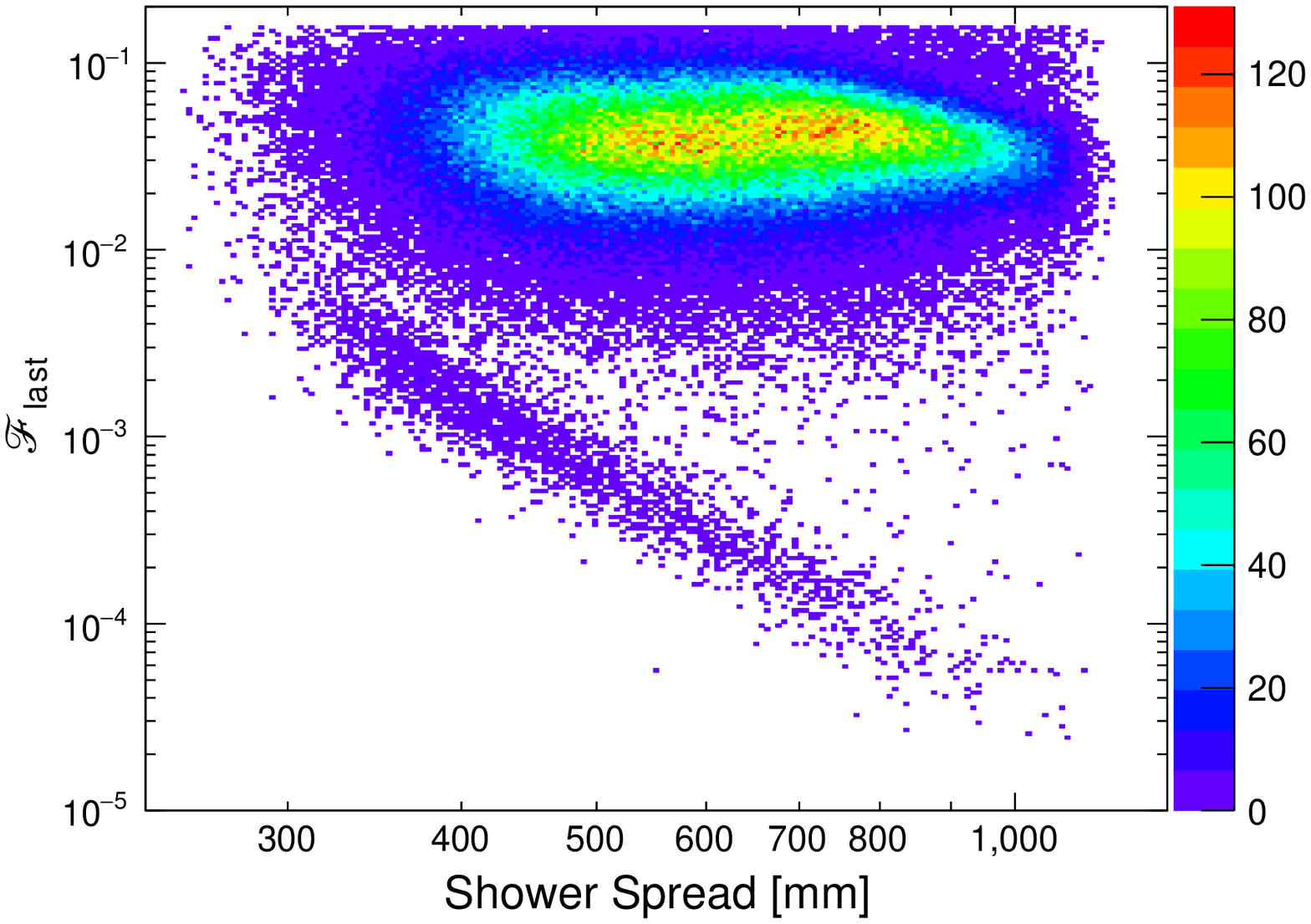}
\includegraphics[width=0.48\textwidth]{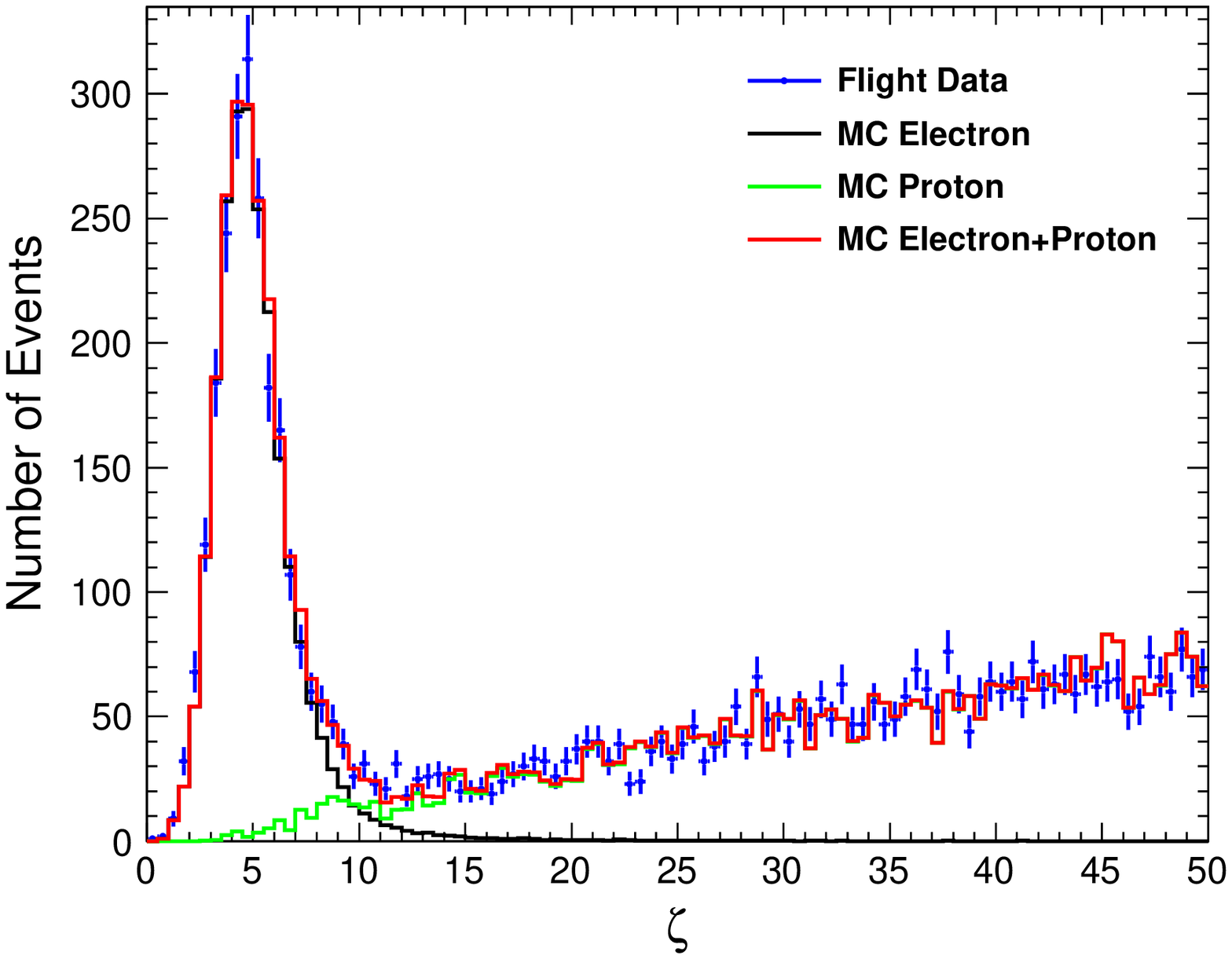}
\caption{Left: Shower spread ($\sum_i{\rm RMS}_i$) versue 
${\mathcal F}_{\rm last}$ for selected events with BGO energies between
500 GeV and 1 TeV. 
Right: one-dimensional distributions of the shower shape parameter $\zeta$,
compared with MC simulations. Plot from Ref. \cite{2017Natur.552...63D}.
\label{fig:ep}}
\end{figure*}

Using 530 days of DAMPE data, the first measurement of the CRE spectrum
from 25 GeV to 4.6 TeV was given \cite{2017Natur.552...63D}. We briefly
describe the data analysis here. The PSD of DAMPE can effective reject 
heavy nuclei and $\gamma$-rays, leaving mainly CREs and protons. 
The shower images in the BGO calorimeter are then used to distinguish 
CREs from protons \cite{1999ICRC....5...37C,2008AdSpR..42..431C}. 
The events with reconstructed tracks passing through the whole detector 
were selected. An energy-weighted root-mean-square (RMS) value of hit
positions in the calorimeter is employed to describe the transverse
spread of a shower. The RMS value of the $i$th layer is defined as
\begin{equation}
{\rm RMS}_i=\sqrt{\frac{\sum_j (x_{j,i}-x_{c,i})^2E_{j,i}}{\sum_j E_{j,i}}},
\end{equation}
where $x_{j,i}$ and $E_{j,i}$ are coordinate and deposit energy of the 
$j$th bar in the $i$th layer, and $x_{c,i}$ is the coordinate of the 
shower center of the $i$th layer. The second parameter, 
${\mathcal F}_{\rm last}$, the deposit energy fraction of the last BGO
layer, is to describe the tail behavior of a shower. The left panel of
Fig.~\ref{fig:ep} shows the scattering plot of the total RMS values of
14 layers ($\sum_i{\rm RMS}_i$) versus ${\mathcal F}_{\rm last}$, for 
events with deposit energies between 500 GeV and 1 TeV. We can clearly
see two populations of the data, with the upper one being proton 
candidates and the lower one being CRE candidates.

To quantify the selection of CREs, these two-dimensional parameter 
distributions were prejected to one-dimension, with a re-defined parameter
$\zeta={\mathcal F}_{\rm last}\times(\sum_i{\rm RMS}_i/{\rm mm})^4/
(8\times10^6)$. The distributions of $\zeta$ for the flight data and
MC simulations of electrons and protons are shown in the right panel
of Fig.~\ref{fig:ep}. A very good match between the data and simulations
can be seen. The CREs are selected with $\zeta \leq 8.5$, which results
in $\sim2\%$ contamination of the proton background based on a fit with
MC templates \cite{2017Natur.552...63D}.

\begin{figure*}[!htb]
\centering
\includegraphics[width=0.48\textwidth]{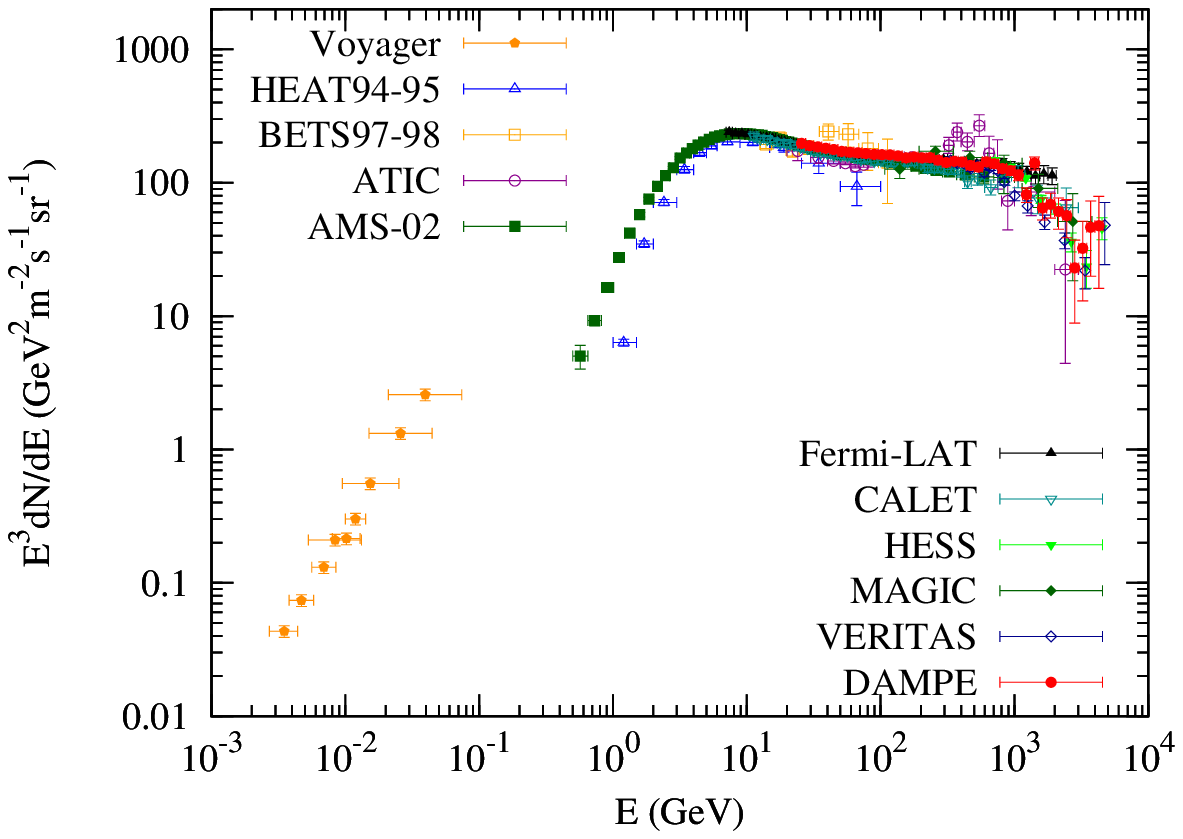}
\includegraphics[width=0.48\textwidth]{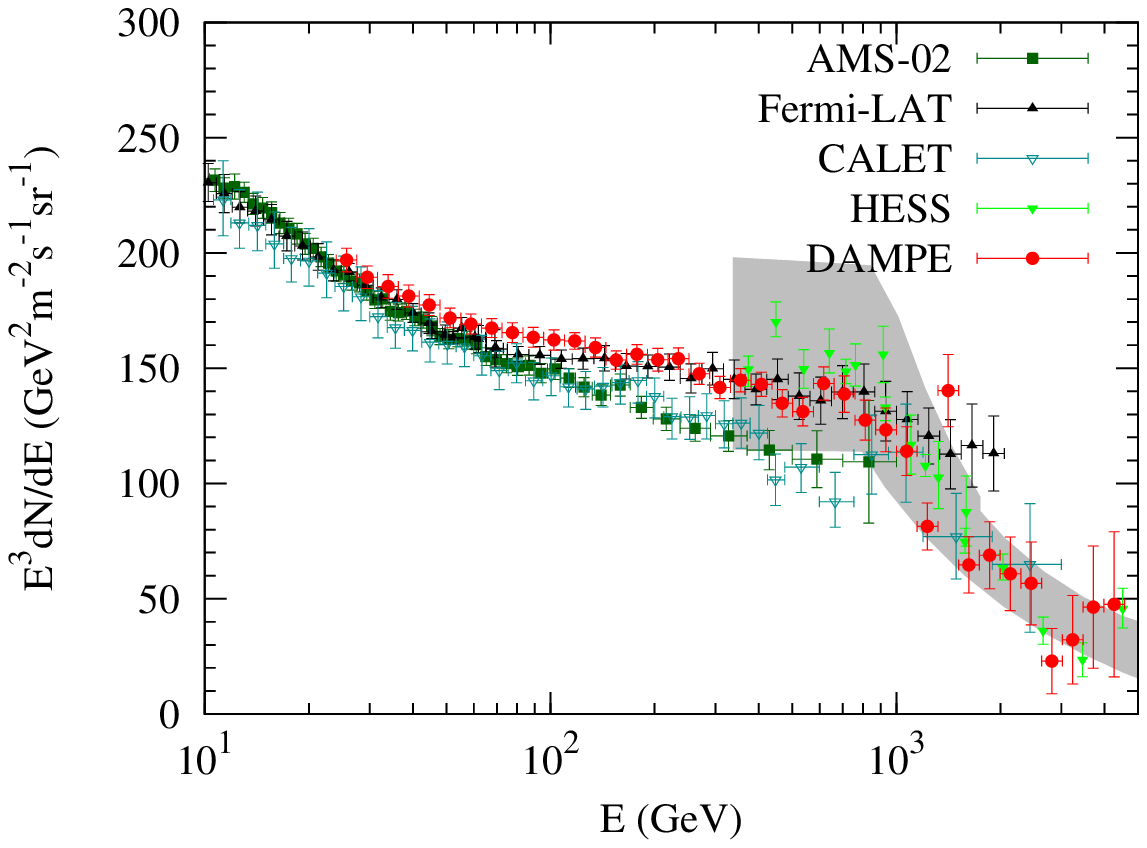}
\caption{CRE spectra measured by Voyager \cite{2016ApJ...831...18C},
HEAT \cite{2001ApJ...559..296D}, BETS \cite{2001ApJ...559..973T},
ATIC \cite{2008Natur.456..362C}, AMS-02 \cite{2014PhRvL.113v1102A},
Fermi-LAT \cite{2017PhRvD..95h2007A}, CALET \cite{2017PhRvL.119r1101A},
DAMPE \cite{2017Natur.552...63D}, and ground-based experiments 
HESS \cite{2008PhRvL.101z1104A,2009A&A...508..561A}, 
MAGIC \cite{2011ICRC....6...47B}, VERITAS \cite{2015ICRC...34..411S}. 
Shaded band in the right panel represents the systematic uncertainties 
of the HESS measurements.
\label{fig:electron}}
\end{figure*}

In total 1.5 million CREs with energies above 25 GeV are selected in
the 2.8 billion events of the data sample. The derived CRE spectrum 
from 25 GeV to 4.6 TeV \cite{2017Natur.552...63D} is shown in 
Fig.~\ref{fig:electron}, compared with other measurements. To highlight 
the high-energy part, we zoom in and show again the spectra above $10$ 
GeV in the right panel of Fig.~\ref{fig:electron}. We can see that 
the DAMPE data significantly reduces the uncertainties of the 
measurement for $E\gtrsim 500$ GeV. The direct measurement of the CRE 
spectrum has been extended to $\sim5$ TeV for the first time. 
Previously, only the IACTs on the ground can reach such high energies, 
but with quite large systematic uncertainties. The overall behavior 
of the DAMPE CRE spectrum is consistent with that of Fermi-LAT and HESS. 
However, for $E\gtrsim100$ GeV, the DAMPE CRE spectrum is slightly 
harder than the results of AMS-02 and CALET. The uncertainties of the 
absolute energy scale of these measurements may partially account for 
such a disagreement.

The DAMPE spectrum clearly reveals a spectral break at $\sim 0.9$ TeV, 
with the spectral index changing from $-3.1$ to $-3.9$. The significance
of such a break is estimated to be about $6.6\sigma$. The spectral indices
are quite consistent with that of Fermi-LAT below the break and that of
IACTs above the break. Previous measurements by IACTs showed only weak
evidence for such a spectral break \cite{2008PhRvL.101z1104A}. The Fermi-LAT
measurement up to 2 TeV and CALET measurement up to 3 TeV, however, did
not show significant spectral break around TeV. This may be due to the
relatively high proton background and low energy resolution of Fermi-LAT
and low statistics of CALET. There is also an indication of a narrow 
spectral peak at energies of $\sim1.4$ TeV. 

Apart from the spectral break around TeV, the CRE spectra at lower 
energies are not featureless either. The Voyager spacecraft has 
travelled for more than 140 astronomical units (AU) away from the 
earth in about 40 years since 1977, and there was evidence showing 
that it passed through the heliopause and entered the interstellar 
space \cite{2013Sci...341..150S}. The CRE spectrum measured by Voyager 
thus represents the local interstellar (LIS) result without the solar 
modulation. The other measurements were obtained near the earth, and 
the low energy parts of the spectra ($E\lesssim30$ GeV) are expected 
to be suppressed by the solar magnetic field. There is a break of the 
CRE spectrum around a few GeV, with spectral indices changing from about 
$-0.7$\footnote{This value depends on solar activities during the 
measurement.} to $-3.2$ \cite{2014PhRvL.113v1102A}. The Fermi-LAT data 
reveals a second break (hardening) at $E\sim50$ GeV, and the spectral 
index changes from $-3.2$ to $-3.1$ \cite{2017PhRvD..95h2007A}. 
We leave the interpretation of the CRE spectral behaviors in Sec. IV.

\section{Physical implications}

\subsection{CRE propagation in the Milky Way}

Before discussing the physical modeling of the CRE spectrum, we briefly
introduce the propagation of CREs. The propagation of CREs in the Milky
Way can be described by a diffusion-cooling eauqtion
\begin{equation}
\frac{\partial \psi}{\partial t}=\nabla\cdot(D\nabla\psi)+
\frac{\partial}{\partial E}(b\psi)+q_e,\label{eq:prop-e}
\end{equation}
where $D=D(E)$ is the diffusion coefficient, $b=b(E)=-{\rm d}E/{\rm d}t$ 
is the cooling rate, and $q_e({\boldsymbol x},E,t)$ is the source injection 
term. This equation is valid at high energies (e.g., $E\gtrsim 10$ GeV) 
where cooling is important. At low energies other physical processes
such as reacceleration or convection may be important
\cite{2009A&A...501..821D}. 

\begin{figure*}[!htb]
\centering
\includegraphics[width=0.48\textwidth]{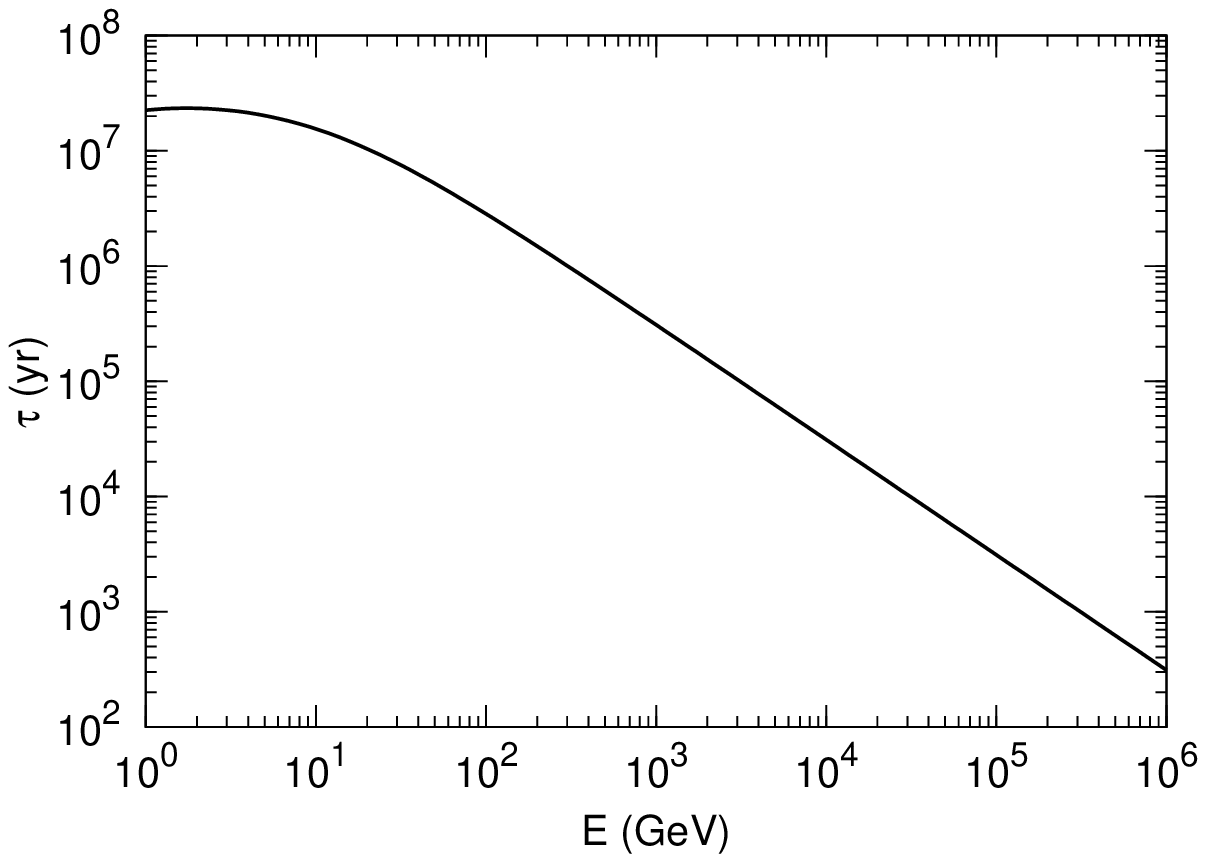}
\includegraphics[width=0.48\textwidth]{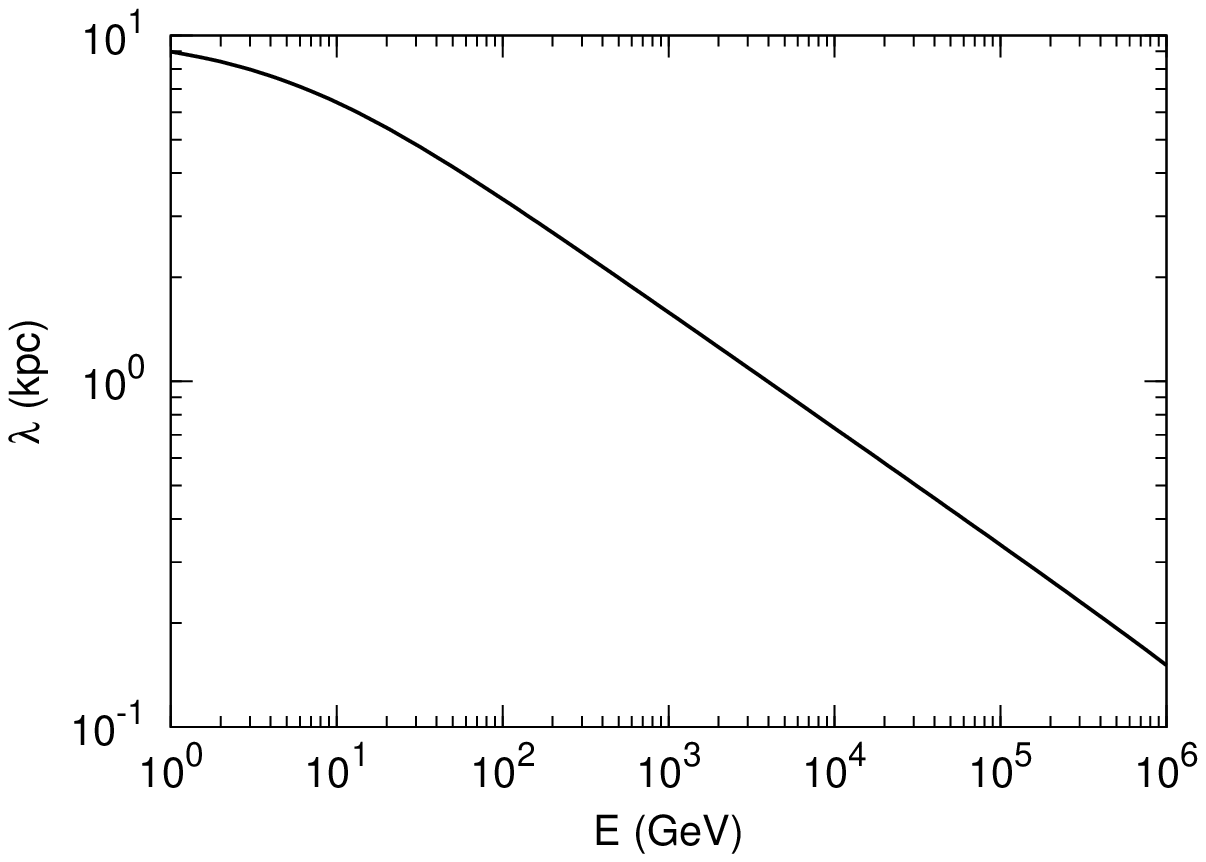}
\caption{Cooling time (left) and effective propagation distance of CREs
as functions of energy.
\label{fig:tl}}
\end{figure*}

Assuming a spherically symmetric geometry with infinite boundary conditions,
the Green's function of Eq. (\ref{eq:prop-e}) with respect to $r$ and $t$
(for $\delta$-type source function in space and time) is
\cite{1995PhRvD..52.3265A}
\begin{equation}
{\mathcal G}(r,E,t)=\frac{N_{\rm inj}(E_i)b(E_i)}{\pi^{3/2}b(E)\lambda^3}
\exp\left(-\frac{r^2}{\lambda^2}\right). \label{eq:propagator}
\end{equation}
Here $q_e=\delta(t)\delta({\bf x})N_{\rm inj}(E)$ with $N_{\rm inj}$ the
injection energy spectrum of electrons, $E_i$ is the initial energy of an
electron which is cooled down to $E$ within time $t$, and
\begin{equation}
\lambda(E)=2\left(\int_E^{\infty}\frac{D(E')}{b(E')}dE'\right)^{1/2},
\end{equation}
is the effective propagation distance of an electron before getting cooled.
The convolution of Eq. (\ref{eq:propagator}) with the source spatial 
distribution and injection history gives the propagated electron fluxes 
at the Earth's location. Fig.~\ref{fig:tl} gives the cooling time 
($\tau(E)\equiv E/b(E)$) and effective propagation distance (right) of
CREs, for typical diffusion and cooling parameters (see Ref.
\cite{2017arXiv171110989Y}). We find that, for TeV electrons, the cooling
time is only about Myr and the propagation distance is about kpc, which
means that TeV electrons can only come from nearby and fresh sources.

\begin{figure*}[!htb]
\centering
\includegraphics[width=0.48\textwidth]{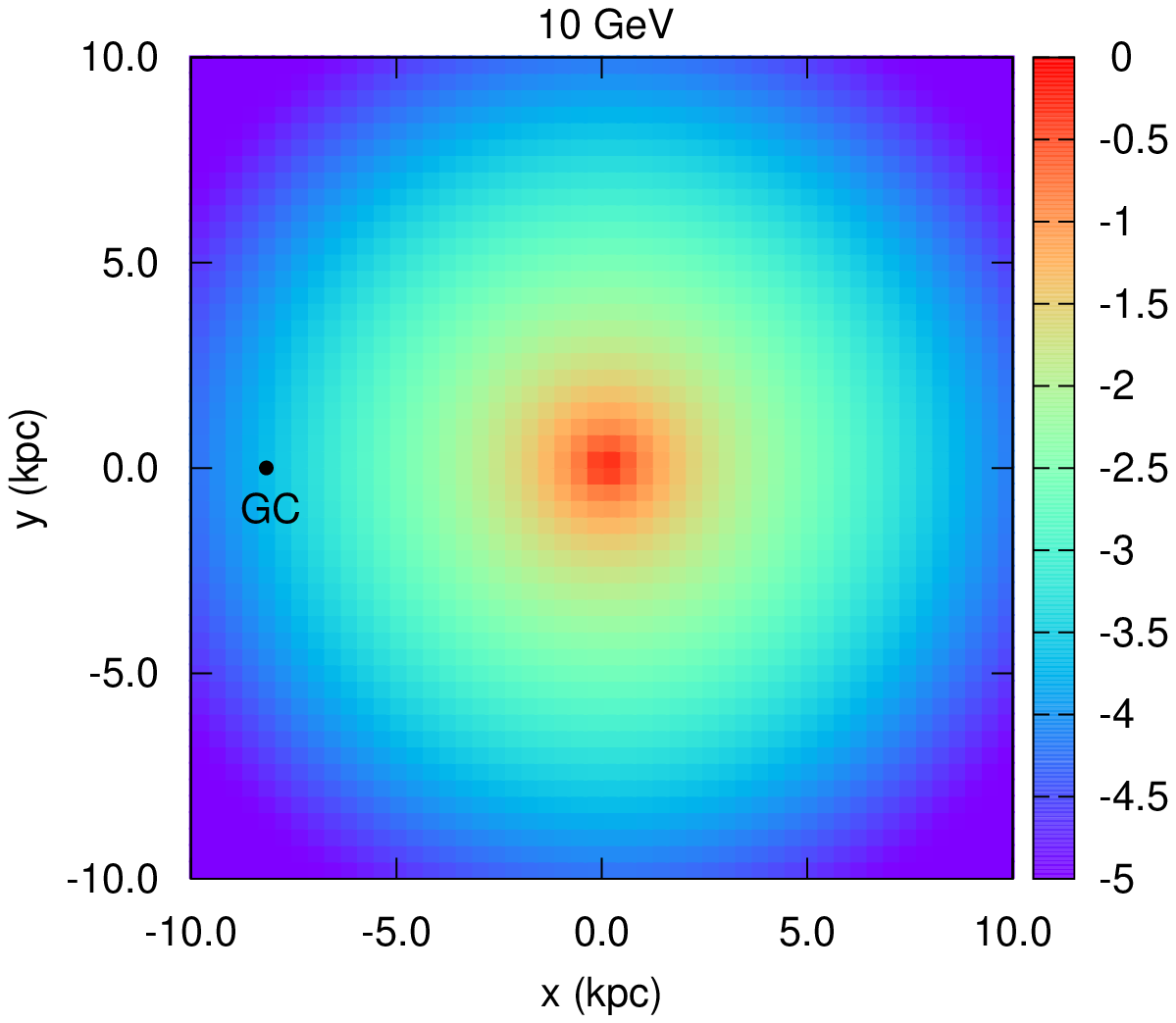}
\includegraphics[width=0.48\textwidth]{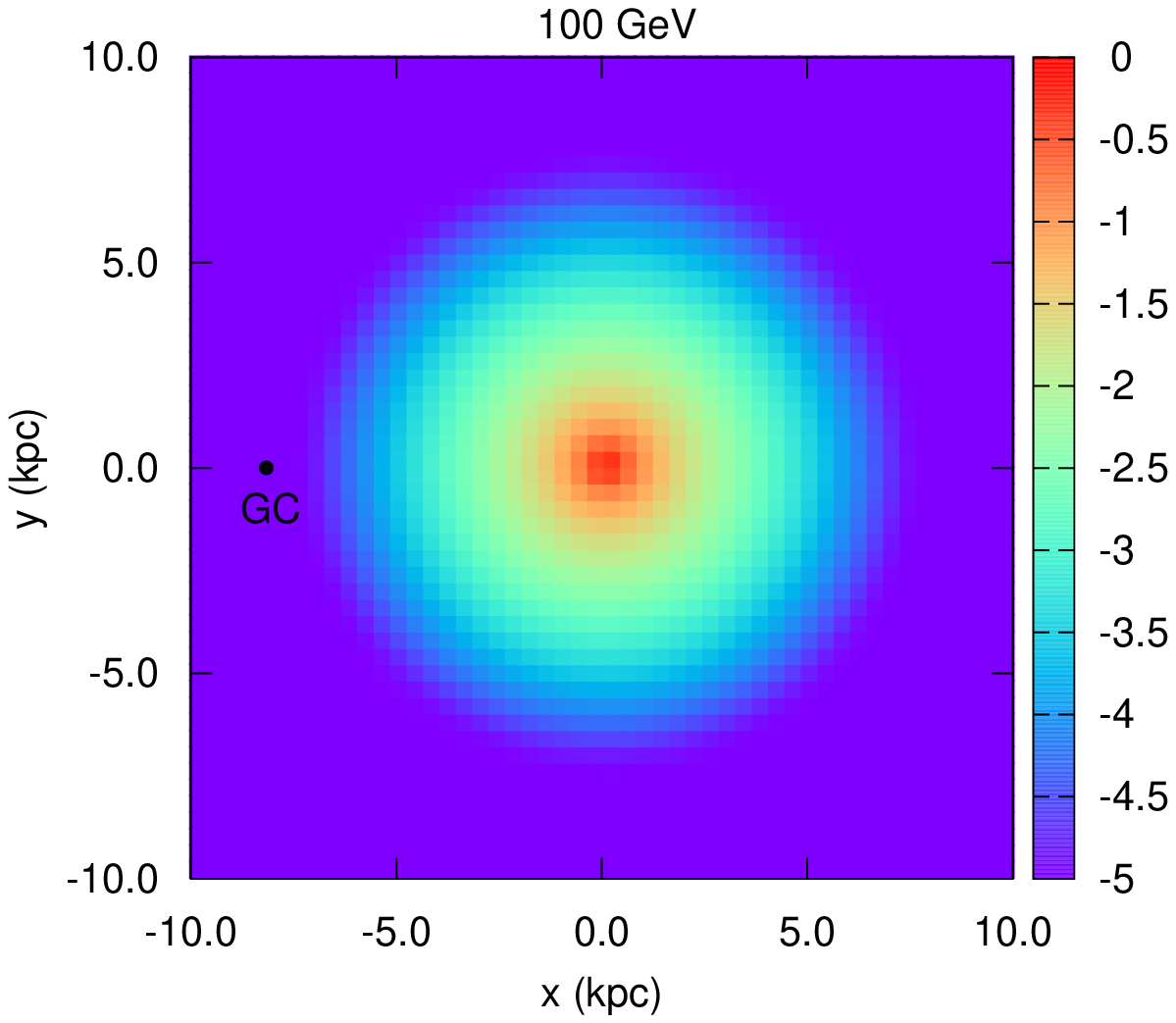}
\includegraphics[width=0.48\textwidth]{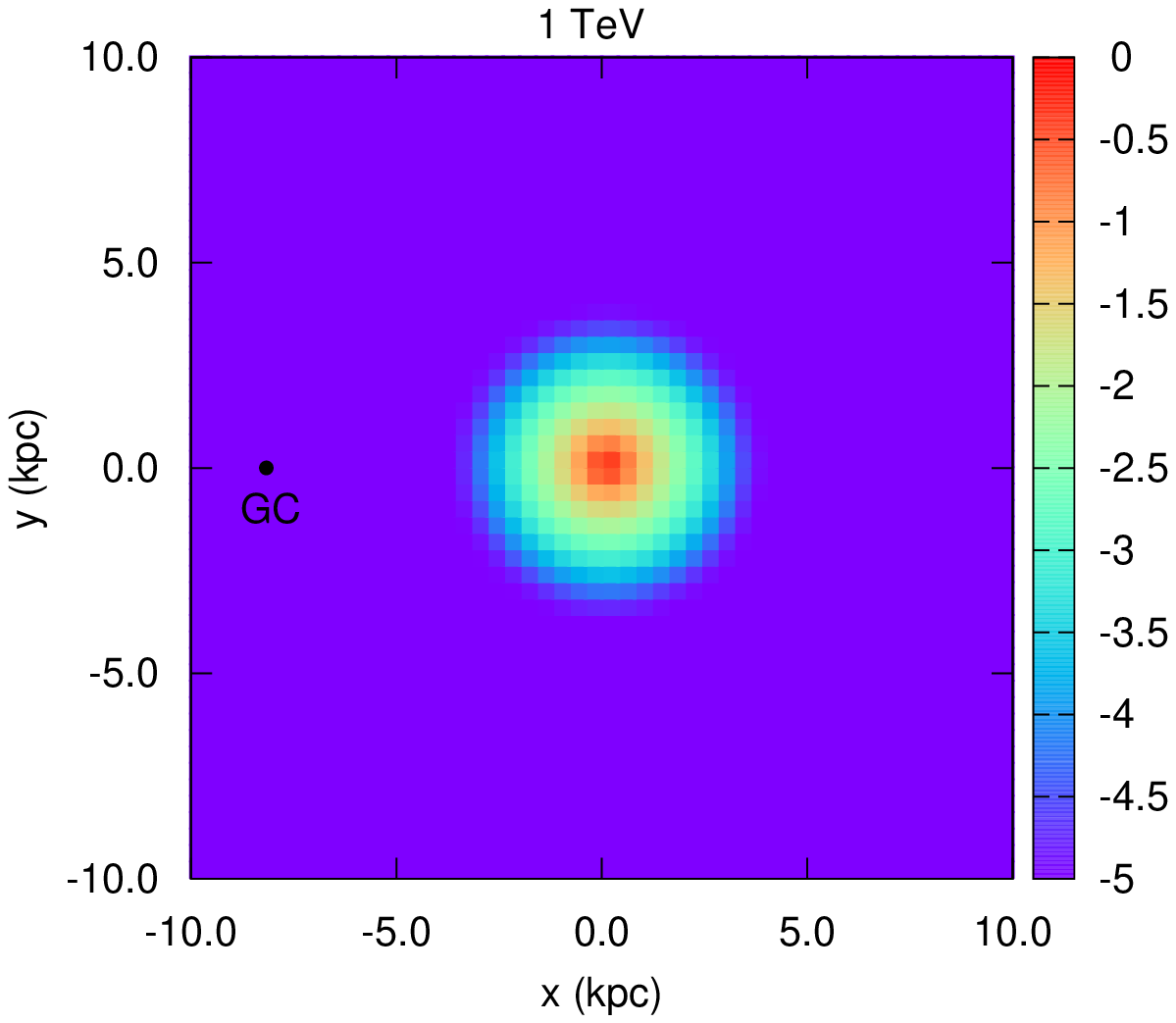}
\includegraphics[width=0.48\textwidth]{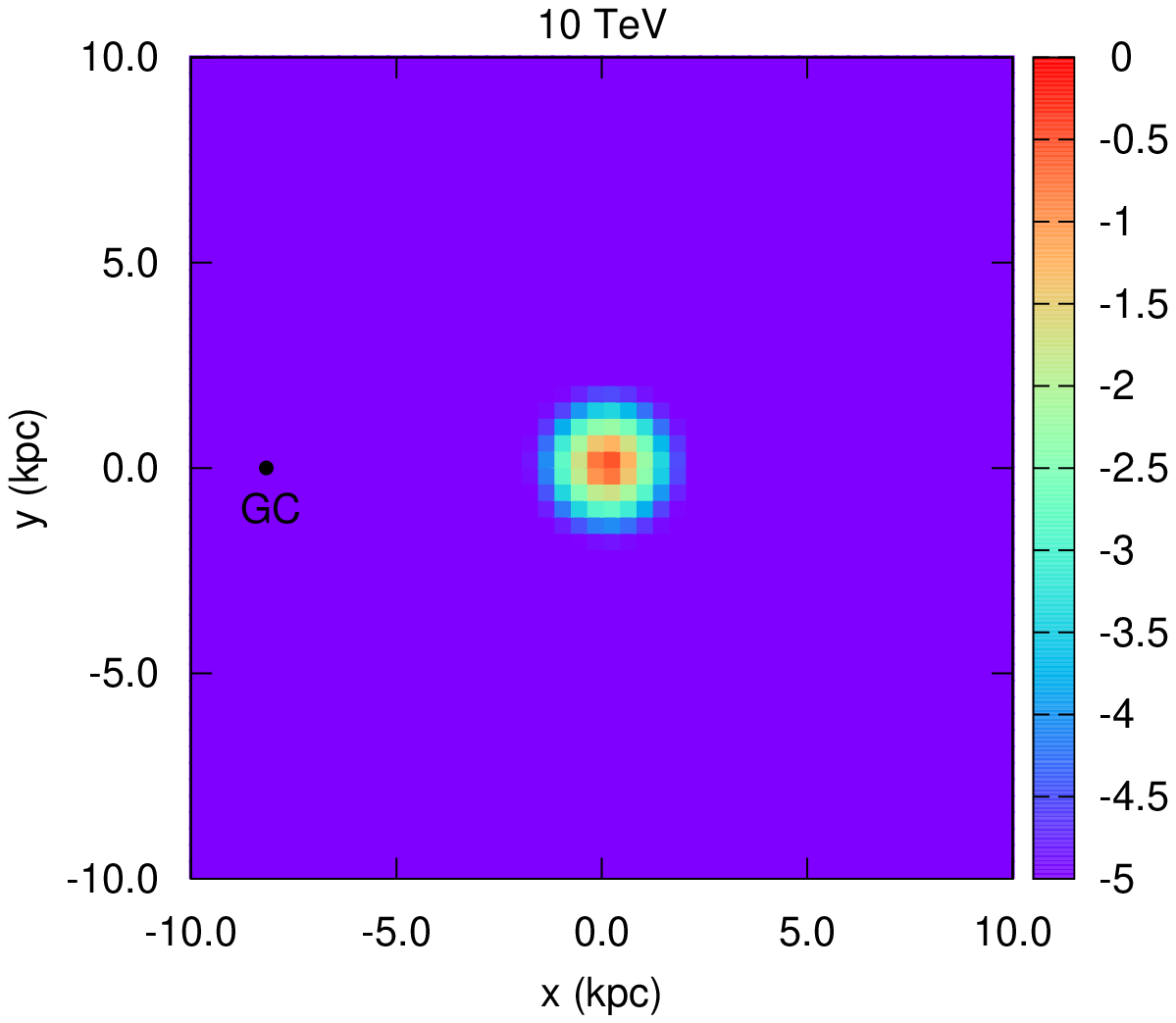}
\caption{Relative logarithmic probability of observing an electron with 
different energy from different location in the Milky Way.
\label{fig:prob}}
\end{figure*}

Fig.~\ref{fig:prob} illustrates the relative (logarithmic) probability 
of observing an electron with specified energy from a source located at 
different distance from the Earth (image center), assuming a continuous 
injection of an $E^{-2}$ power-law. The $xy$ plane refers to the Galactic 
plane, and the $z$ direction points to the Galactic poles. This plot
shows again that high energy CREs should have a local origin.

\subsection{Overall modeling of the CRE spectrum and the positron excess}

The CREs defined in this review include both electrons and positrons.
It has been widely accepted that these CREs consist of several components:
\begin{itemize}

\item {\it Primary electrons} which are accelerated in accompany with 
nuclei CRs by Galactic acceleration sources, e.g., supernova remnants.

\item {\it Secondary electrons and positrons} from the inelastic
collisions between CR nuclei (mostly protons and Helium nuclei) and
the interstellar medium (ISM).

\item {\it Additional electrons and positrons} from some kind of yet
unknown sources which are employed to explain the positron excess
\cite{2009Natur.458..607A,2012PhRvL.108a1103A,2013PhRvL.110n1102A}.

\end{itemize}

For energies from 0.5 GeV to $\sim 500$ GeV, the positron fraction in
CREs is about $(5 \sim 20)\%$ \cite{2014PhRvL.113l1101A}. Thus the CRE
flux is dominated by {\it primary electrons} in such an energy range.
At higher energies, the behavior of the positron fraction is not clear
yet, and whether the {\it primary electrons} can still dominate or not
is not sure.

The overall CRE spectrum and positron fraction can be understood in 
such a three-component scenario. The first spectral break at several
GeV is due to several effects. The ionization and Coulomb cooling
plays a significant role in regulating the low-energy ($<$GeV) spectrum 
of CREs \cite{1998ApJ...493..694M}. The fit to the data further suggests
a break of the injection spectrum at a few GeV \cite{2012PhRvD..85d3507L,
2015APh....60....1Y}. The observations of $\gamma$-ray emission from 
supernova remnants did suggest a broken power-law form of the particle
spectrum \cite{2013Sci...339..807A}. The underlying physics of such an 
injection break is not clear yet. It was proposed that strong ion-neutral 
collisions near the shock fronts of CR acceleration sources 
\cite{2011NatCo...2E.194M} or the escape of particles from/into 
finite-size region might lead to this break \cite{2011MNRAS.410.1577O,
2010MNRAS.409L..35L}. The solar modulation suppresses the low-energy 
flux more than the high-energy flux, and also contribute to form a break. 
All these effects together give the observed spectral break at a few GeV.

\begin{figure*}[!htb]
\centering
\includegraphics[width=0.48\textwidth]{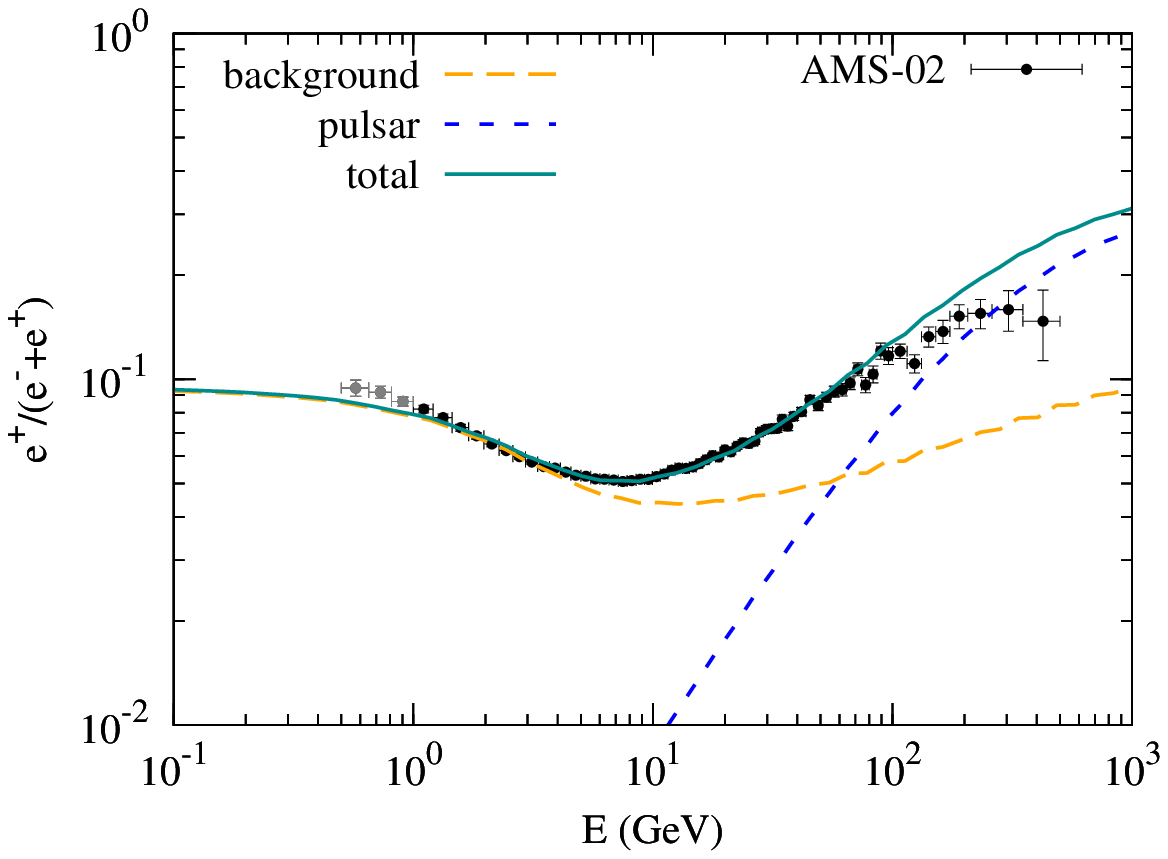}
\includegraphics[width=0.48\textwidth]{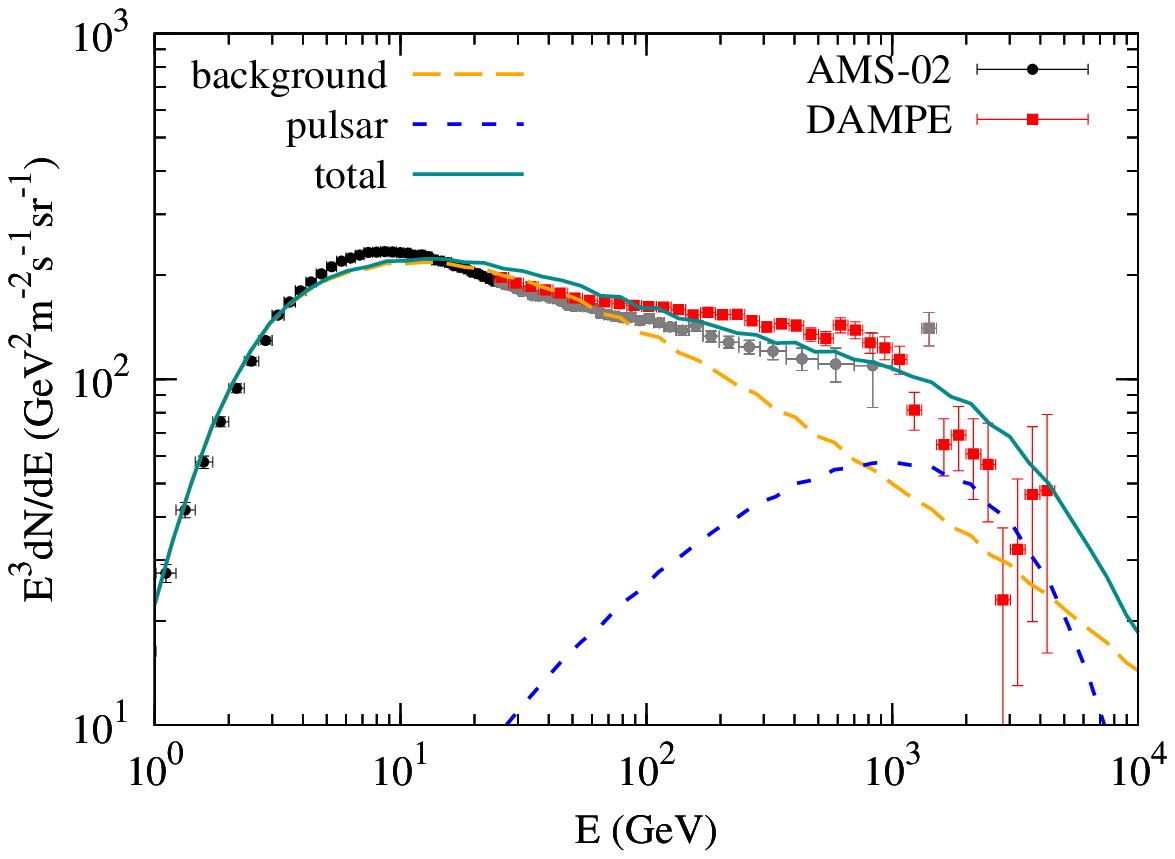}
\includegraphics[width=0.48\textwidth]{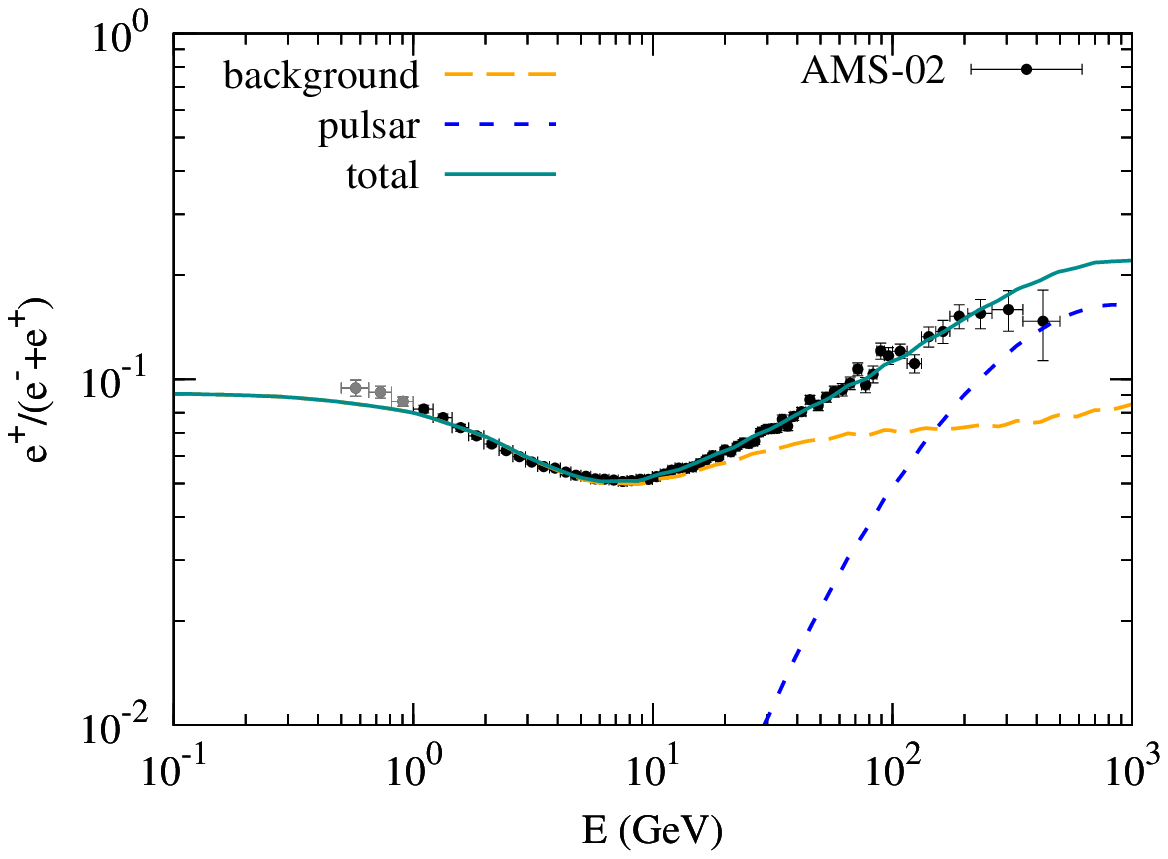}
\includegraphics[width=0.48\textwidth]{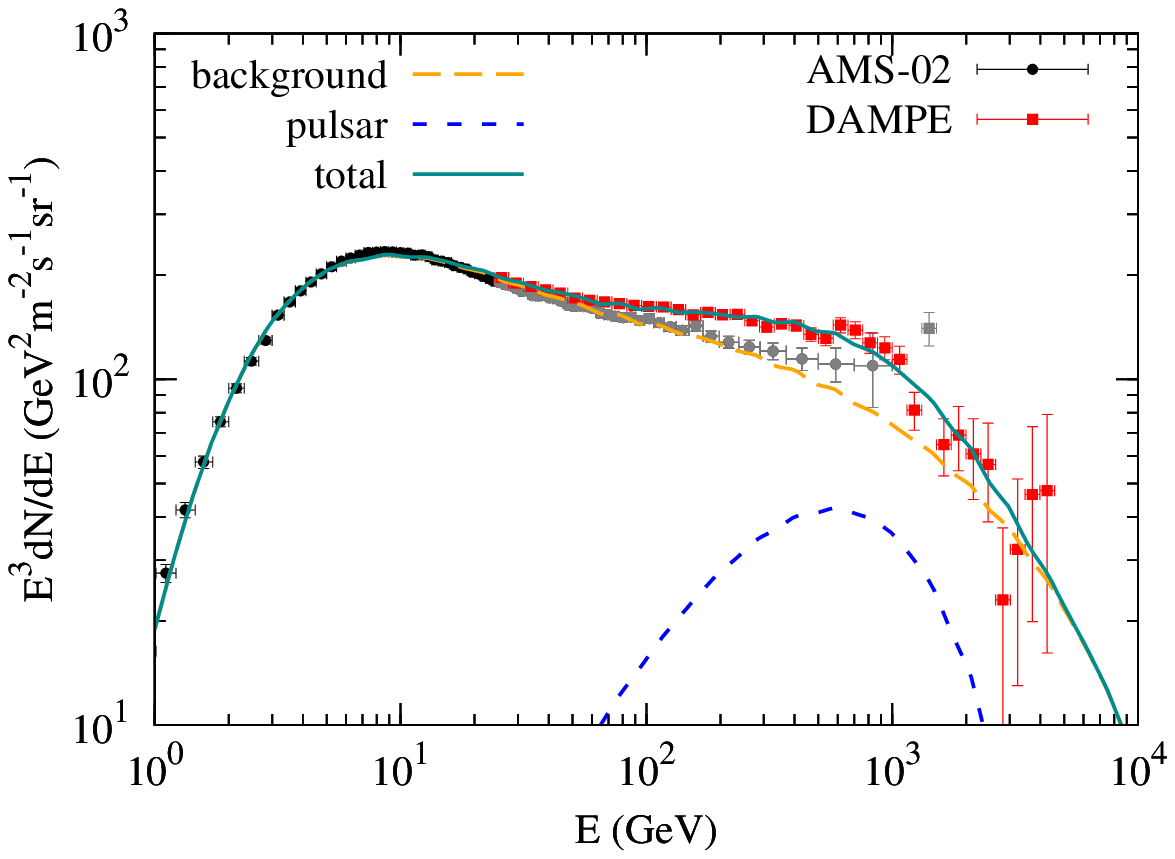}
\caption{Fitting results of the positron fraction and CRE spectra from 
the three component model, assuming that pulsars are {\it additional} 
sources of electrons and positrons. The upper (lower) panels are for the 
model without (with) the spectral hardening of the {\it primary electrons}.
The observational data are AMS-02 positron fraction 
\cite{2013PhRvL.110n1102A}, AMS-02 and DAMPE CRE spectra
\cite{2014PhRvL.113v1102A,2017Natur.552...63D}.
Plot from Ref. \cite{2017arXiv171110989Y}. 
\label{fig:harden}}
\end{figure*}

Quantitative fitting to the positron fraction and CRE spectra with those
three components, assuming either pulsars or DM annihilation/decay as the 
source(s) of the {\it additional electrons and positrons}, shows that
a spectral hardening around $50\sim100$ GeV of the {\it primary electron} 
component is required \cite{2013PhLB..727....1Y,2015JCAP...03..033Y,
2015PhRvD..91f3508L} (see also \cite{2014PhLB..728..250F,2013PhRvD..88b3013C}).
Fig.~\ref{fig:harden} shows the comparison of the fitting results for the
model without (top two panels) and with (bottom two panels) the spectral
hardening of the {\it primary electrons}. The fit can be improved
significantly when including the spectral hardening. The origin of the
spectral hardening, as well as the TeV break revealed by HESS/DAMPE 
(discussed in more details below), may be due to the breakdown of 
continuous source distribution and imprints of nearby source(s)
\cite{1995A&A...294L..41A,2014JCAP...04..006D,2017ApJ...836..172F}.

\subsection{Implication of the TeV spectral break}

\begin{figure*}[!htb]
\centering
\includegraphics[width=0.48\textwidth]{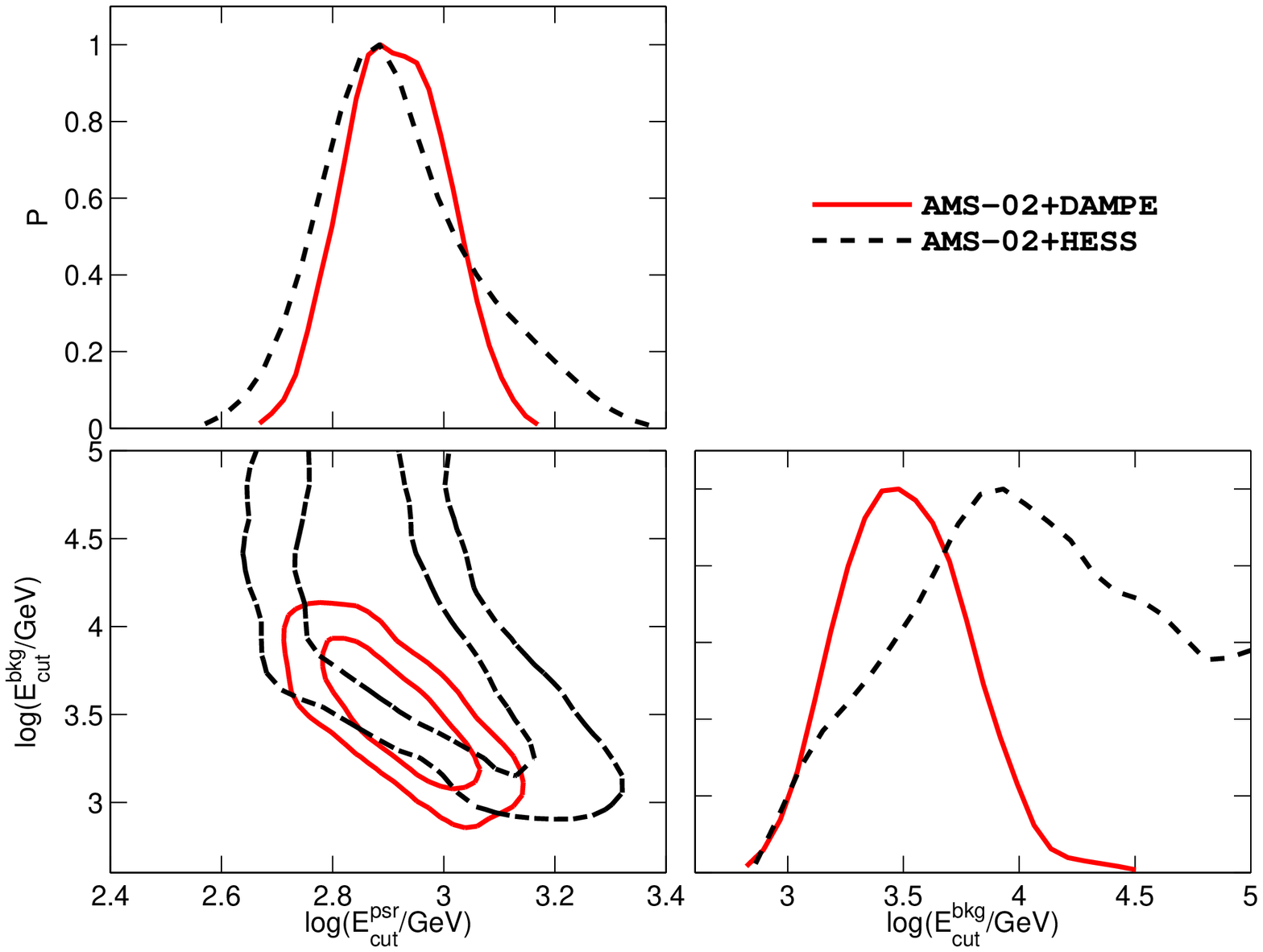}
\includegraphics[width=0.48\textwidth]{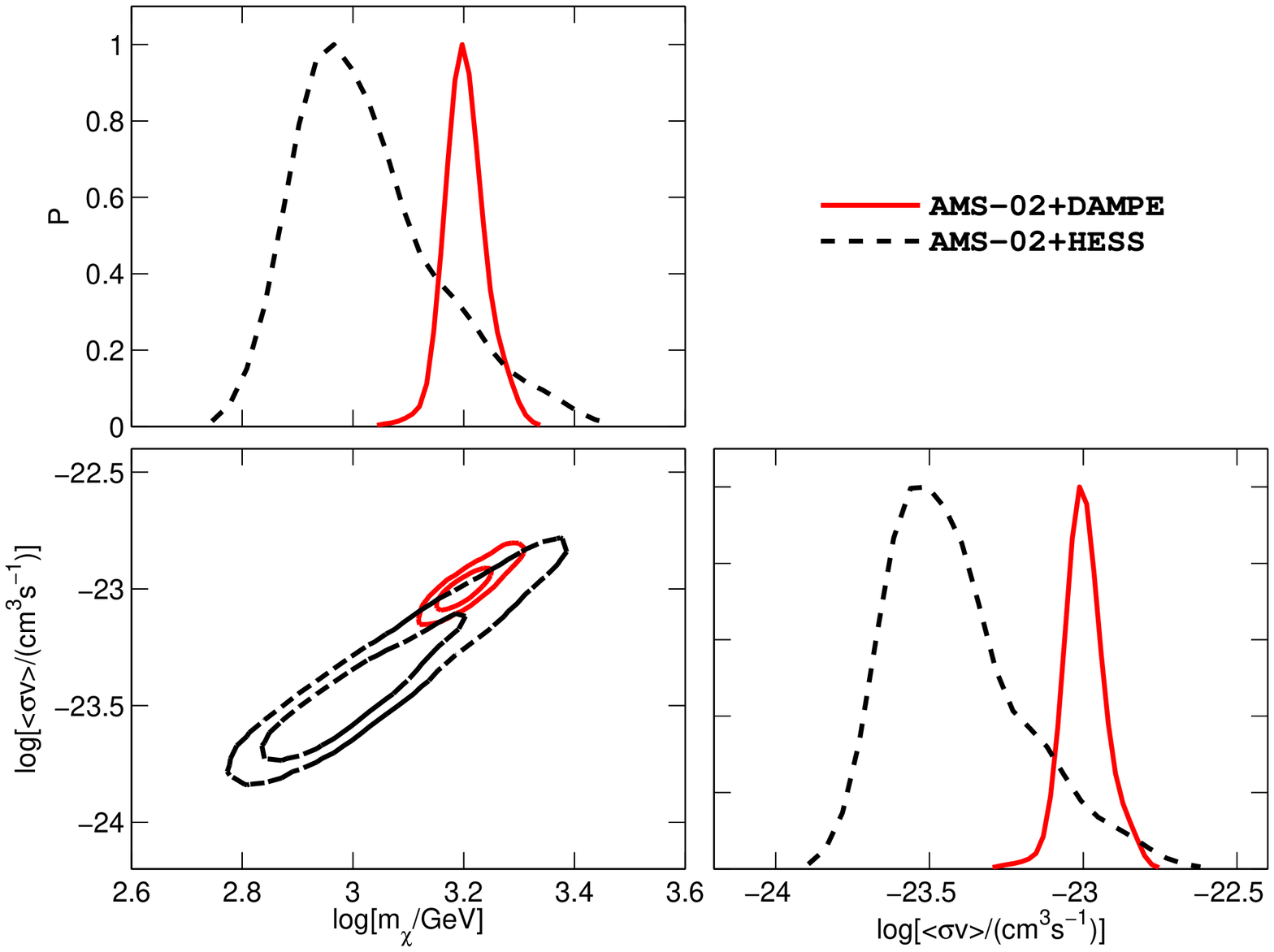}
\caption{Left: constraints on the cutoff energies of the {\it primary}
electrons and that from pulsars. Right: constraints on the DM mass and 
annihilation cross section assuming $\mu^+\mu^-$ channel. In each panel, 
the bottom-left sub-panel shows the 68\% and 95\% confidence level 
contours, and the diagonal sub-panels show the one-dimensional
probability distributions. Black dashed lines are for the fitting to
pre-DAMPE data, and red solid lines are for the AMS-02 + DAMPE data.
Plot from Ref. \cite{2017arXiv171110989Y}.
\label{fig:triangle}}
\end{figure*}

\begin{figure*}[!htb]
\includegraphics[width=0.48\textwidth]{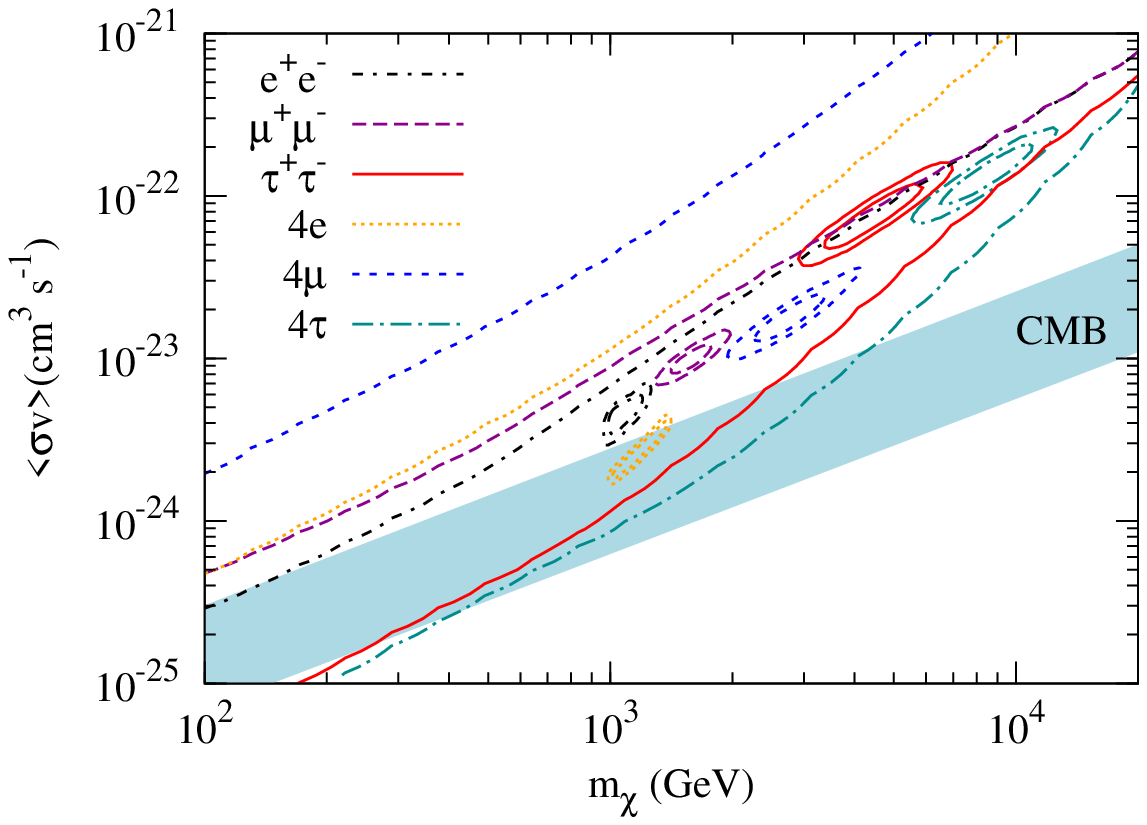}
\includegraphics[width=0.48\textwidth]{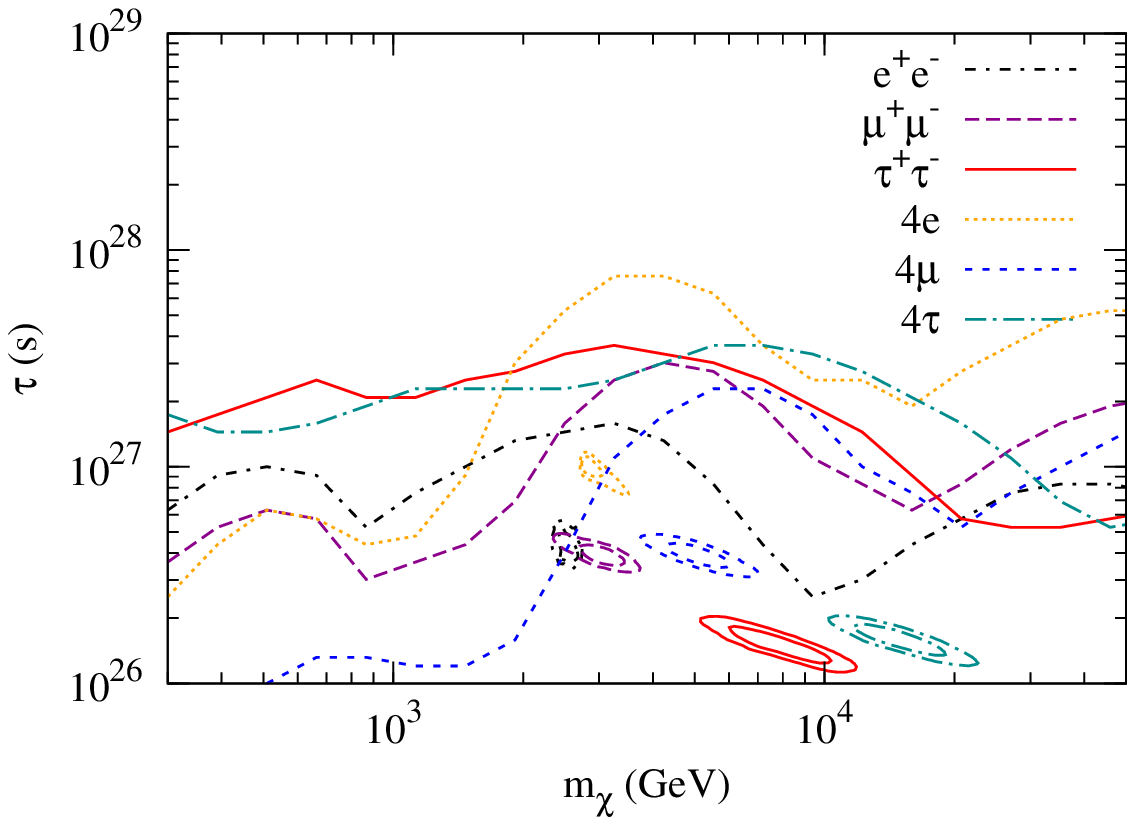}
\caption{Favored parameter regions on the $m_{\chi}-\sv$ (for annihilation;
left) and $m_{\chi}-\tau$ (for decay; right) plane to explain the AMS-02
positron fraction and DAMPE CRE spectrum, for channels $e^+e^-$, $\mu^+\mu^-$, 
$\tau^+\tau^-$, $4e$, $4\mu$, and $4\tau$, respectively. Also shown are
the constraints from Planck observations of CMB and Fermi-LAT observations
of dwarf galaxies \cite{2017CoPhC.213..252H} (left) and the extragalactic 
$\gamma$-ray background \cite{2015ApJ...799...86A} (right). 
Plot from Ref. \cite{2017arXiv171110989Y}.
\label{fig:con}}
\end{figure*}

The spectral break around TeV as observed by IACTs and DAMPE has very
interesting physical implications. The precise CRE spectrum by DAMPE
can significantly narrow down the parameter space of models, either
astrophysical ones or DM annihilation/decay, to explain the positron 
excess \cite{2017arXiv171110989Y}. Fig.~\ref{fig:triangle} shows the
constraints on some parameters of the pulsar model (left) and the
DM annihilation model (right). To give such constraints, we again
assume the three-component model as described above, and add an
exponential cutoff with characteristic energy $E_{\rm cut}^{\rm bkg}$ 
of the {\it primary} electron spectrum. The injection spectrum of 
electrons and positrons from pulsars is parameterized as an exponential 
cutoff power-law form, with cutoff energy $E_{\rm cut}^{\rm psr}$.
For the DM annihilation model, the density profile is assumed to be
a Navarro-Frenk-White form \cite{1997ApJ...490..493N}, and the 
annihilation channel is $\chi\chi\to\mu^+\mu^-$. Two results are
compared: the fitting to pre-DAMPE CRE data (i.e., AMS-02 and HESS)
and to the DAMPE data. In both fittings the AMS-02 positron fraction
is included. It shows that the DAMPE data reduce the parameter space
remarkably. Most importantly, improved constraints on the model
parameters lead to robust exclusion of the simple DM annihilation or 
decay models to account for the positron excess, when considering
the constraints from the cosmic microwave background (CMB; 
\cite{2016A&A...594A..13P}) and Fermi-LAT $\gamma$-ray observations 
\cite{2017CoPhC.213..252H,2015ApJ...799...86A}, as shown in 
Fig.~\ref{fig:con}. Note that in Ref. \cite{2017arXiv171110989Y}
the fittings for the $e^+e^-$ channel of the DM models are poor with
too large $\chi^2$ values. Therefore the corresponding contours in
Fig.~\ref{fig:con} for such a channel are statistically less meaningful.
For other channels the fittings are acceptable.

The TeV break could be naturally explained by one or a few nearby 
source(s). The cooling effect during the propagation of CREs could
form a cutoff of the detected spectrum (e.g., \cite{1995PhRvD..52.3265A,
1995A&A...294L..41A}). This ``knee-like'' structure might also be due
to the maximum energy of CREs injected into the Milky Way. In Ref. 
\cite{2018ApJ...854...57F} it was proposed that the confinement and 
radiative cooling of electrons during the acceleration in supernova 
remnants could account for this CRE knee. Such electrons escape from 
the sources after $\sim10^5$ yr when the supernova remnants merge into 
the ISM, and the resulting CRE spectrum naturally cuts off around TeV 
\cite{2007A&A...465..695Z}. This scenario may work. We just emphasize 
the importance of discretness of the source distribution for understanding 
the TeV spectrum of CREs. Assuming a supernova rate of $10^{-2}$ per
year, the estimated number of supernovae within the cooling time and
the effective propagation distance is $O(10)$ for TeV electrons.
Therefore the active sources of TeV CREs should be discrete.

\subsection{Discussion on the tentative 1.4 TeV structure}

The tentative peak structure at $\sim 1.4$ TeV, if gets confirmed with
more data, is completely unexpected. We emphasize that the current data
is not significant enough to draw a conclusion. The local significance
is estimated to be about $3.7\sigma$ assuming a broken power-law
background \cite{2017arXiv171200005H,2018PhLB..780..181F}. If the
look-elsewhere effect is taken into account, the estimated global
significance is about $2.3\sigma$ \cite{2018PhLB..780..181F}.
Note, however, the significance estimate depends on the background
adopted, as illustrated in Ref. \cite{2017arXiv171202744G}. 
Nevertheless, given the potential importance of such a structure, 
quite a number of works appeared to discuss its possible implications 
\cite{2017arXiv171110989Y,2017arXiv171200005H,
2017arXiv171110995F,2018PhLB..778..292G,2018JHEP...02..107D,
2017arXiv171111182C,2017arXiv171111452C,2017arXiv171111579L,
2017arXiv171111333G,2017arXiv171111563D,2017arXiv171201239G,
2018EPJC...78..198C,2017arXiv171203652O,2018EPJC...78..216H,
2018arXiv180104729N,2018PhLB..779..130L,2017arXiv171200941N,
2017arXiv171203642S,2017arXiv171111052Z,2017arXiv171111058T,
2018JHEP...02..121A,2017arXiv171200037C,2017arXiv171200362J,
2017arXiv171200370G,2017arXiv171200372N,2018PhRvD..97f1302C,
2017arXiv171200922G,2017arXiv171201143Z,2017arXiv171201724Y,
2017arXiv171202021D,2018ChPhC..42c5101L,2017arXiv171202744G,
2017arXiv171203210Z,2017arXiv171205351C,2017arXiv171209586N,
2016arXiv161108384J,2017arXiv171207868J}.

We summarize some general requirements to account for the data.
First is about the energetics. We discuss two kinds of injection modes 
of the sources: instantaneous injection and continuous injection. 
For instantaneous injection, the total energy of the source is required 
to be \cite{2017arXiv171110989Y}
\begin{eqnarray}
\varepsilon_{\rm tot} & \sim & 3 \times 10^{47}~{\rm erg}~\left(\frac{D(E)}
{3 \times 10^{29}~{\rm cm^{2}~s^{-1}}}\right)^{3/2}\left(\frac{\tau_{\rm s}}
{3 \times 10^5~{\rm yr}}\right)^{3/2}\nonumber\\
& \times & \left({w_{\rm e}\over 1.2\times10^{-18}~{\rm erg~cm^{-3}}}\right),
\end{eqnarray}
where $\tau_{\rm s}$ is the age of the source and $w_e$ is the energy
density of the peak. This total energy release ($10^{46} \sim 10^{48}$
erg depending on different injection time) is a little bit smaller
but still comparable to a typical pulsar \cite{2013PhRvD..88b3001Y}. 
For continuous injection, the source luminosity is 
\cite{2017arXiv171110989Y}
\begin{eqnarray}
\dot{Q}&\approx& 2 \times 10^{33}~{\rm erg~s^{-1}}~\left({R\over 0.1~{\rm kpc}}
\right)\left({D(E)\over 3 \times 10^{29}~{\rm cm^{2}~s^{-1}}}\right)\nonumber\\
&\times &\left({w_{\rm e}\over 1.2\times10^{-18}~{\rm erg~cm^{-3}}}\right),
\end{eqnarray}
where $R$ is the source distance. If the DM annihilation is assumed to
be continuous source of those CREs, the average density of DM is about
$0.012$ TeV cm$^{-3}$ for a region of $\sim 0.1$ kpc, $m_{\chi}=1.5$ TeV,
and a thermal annihilation cross section, which is about 30 times higher 
than the canonical local density. This means a local DM subhalo or a density
enhancement is required \cite{2017arXiv171110989Y,2017arXiv171200005H,
2017arXiv171200362J}. To account for the data, the subhalo mass is found 
to be about $10^7 \sim 10^8$ M$_\odot$ or the local density is enhanced 
by a factor of $17 \sim 35$. Such a requirement is extreme according to
numerical simulations of DM structures, and thus some other mechanisms
are necessary to boost the cross section or density distribution, e.g.,
a mini-spike or an ultra-compact micro halo \cite{2017arXiv171200005H,
2017arXiv171201724Y}.

\begin{figure}[!htb]
\centering
\includegraphics[width=0.48\textwidth]{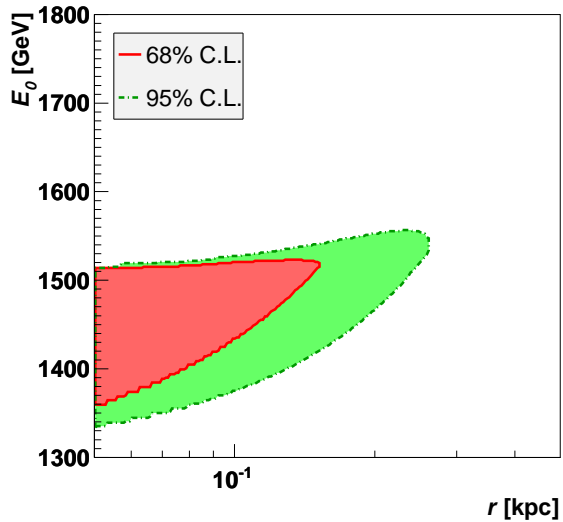}
\caption{Allowed 68\% and 95\% regions for the distance and central energy 
of a continuous point-like source with a Gaussian injection spectrum.
Plot from Ref. \cite{2017arXiv171200005H}.
\label{fig:rE0}}
\end{figure}

As described above, the strong cooling of CREs limited the propagation
distance of an electron before losing most of its energy. For TeV
electrons, such a distance is about 1 kpc \cite{2017arXiv171110989Y}.
The spectral shape gives further constraints on the source distance.
If the distance is too large, the cooling effect would broaden the
CRE spectrum, making it difficult to fit the data. Typically the
distance is required to be smaller than $\sim0.3$ kpc
\cite{2017arXiv171110989Y,2017arXiv171200005H}. Fig.~\ref{fig:rE0}
illustrates the constraints on the distance and injection energy of 
a continuous point-like source with a Gaussian injection spectrum
\cite{2017arXiv171200005H}. In Ref. \cite{2017arXiv171200005H} it
was suggested that for the instantaneous injection scenario, a source
with a distance of $1 \sim 3$ kpc could also give a sharp spectrum.
This is due to that low-energy particles are not able to propagate to 
such a distance, while high-energy electrons have already cooled down. 
A caution is that this scenario may need a quite large energy output 
into electrons/positrons from the source, $10^{48} \sim 10^{51}$ erg. 

The injection spectrum can also not be arbitrary. In general the 
propagation will make the CRE spectrum broader than the injection one. 
Therefore to give a narrow peak around the Earth, the injection spectrum 
should also be quasi-monochromatic. For the DM annihilation/decay, the 
direct $e^+e^-$ channel is required. Astrophysically it is not easy to 
generate quasi-monochromatic spectra. We found a possible exception is 
the cold, ultra-relativistic $e^+e^-$ plasma wind from pulsars 
\cite{1984ApJ...283..694K,2012Natur.482..507A,2012A&A...547A.114A}. 
The Lorentz factor of the bulk flow of pulsar winds is suggested to be 
$\sim10^6$, which just corresponds to the energy of the DAMPE peak.
However, there should not be significant energy loss or gain of such 
pulsar wind during their transportation outwards through the pulsar
wind nebula. Otherwise the narrow injection spectrum cannot be
maintained. A good fit to the data with reasonable pulsar model
parameters was given in Ref. ~\cite{2017arXiv171110989Y}. The model
details, as well as the escape from the nebula, need further studies.
The injection spectrum can be softer for a larger distance of the
source \cite{2017arXiv171200005H}. However, it is still quite hard
($\alpha \lesssim 1.3$) compared with canonical prediction of shock
acceleration.

Within the DM annihilation/decay scenario from a local clump, many works 
investigate the potential particle model extension of the standard model. 
A general requirement is that DM particles should couple with electrons 
(leptons) instead of quarks. A simple extension is to add a $U(1)_X$ 
symmetry to the standard model, and the DM annihilate into 
electrons (leptons) via the new gauge boson mediator, e.g., a $Z'$
\cite{2017arXiv171110995F,2018PhLB..778..292G,2018JHEP...02..107D,
2017arXiv171111182C,2017arXiv171111452C,2017arXiv171111579L,
2017arXiv171111333G,2017arXiv171111563D,2017arXiv171201239G,
2018EPJC...78..198C,2017arXiv171203652O,2018EPJC...78..216H,
2018arXiv180104729N}. In the fermion DM case, the Dirac type is more
natural as its cross section is $s$-wave dominant for the $s$-channel 
annihilation process $\chi\bar\chi \to Z'\to f \bar f$. Also several 
bosonic DM models were developed with different Higgs potentials and 
Yukawa terms. Some of such models are realized in different kinds of 
seesaw models \cite{2017arXiv171110995F,2018PhLB..779..130L,
2017arXiv171200941N,2017arXiv171203642S,2018EPJC...78..216H,
2017arXiv171111333G}. Other attempts include the lepton portal DM
model in which DM particles interact with standard model 
particles through mediators with lepton numbers
\cite{2017arXiv171110989Y,2017arXiv171111058T}, or the lepton-flavored
DM model in which DM particles carry lepton numbers themselves
\cite{2017arXiv171200037C}. 
The singlet scalar DM model \cite{2017arXiv171200372N}, 
higgs triplet with vector DM model \cite{2018PhRvD..97f1302C}, 
graviton-mediated DM model \cite{2017arXiv171201143Z}, 
Quasi-degenerate DM model \cite{2017arXiv171200922G}, 
lepton-specific interaction DM model \cite{2018ChPhC..42c5101L}, 
and left-right symmetric scalar DM model \cite{2017arXiv171205351C} 
were also presented.

Contraints on the DM models with direct detection and collider data 
have been discussed in some models (e.g., \cite{2017arXiv171110995F,
2018JHEP...02..107D,2018PhLB..778..292G, 2017arXiv171111058T,
2017arXiv171111182C,2018JHEP...02..121A}). Since in general the DM 
directly or indirectly couples with leptons in these models, the DM-nucleon 
scattering is induced through the lepton-loop and photon mediator. 
The constraints from the current direct detection data are relatively 
weak. It was shown that for some particular types of interactions, 
the direct detection data can exclude a considerable part of the 
parameter space \cite{2018JHEP...02..107D,2017arXiv171200037C,
2017arXiv171111563D,2017arXiv171201239G}. Such interaction can also 
contribute to the lepton-nucleon scattering through the Drell-Yan 
process, which is also significantly lower than the current LHC  
sensitivity \cite{2018JHEP...02..107D}. The constraints from null 
detection of the leptophilic $Z^{\prime}$ bosons \cite{2018JHEP...02..121A, 
2017arXiv171111563D,2017arXiv171201239G,2017arXiv171203652O,
2018arXiv180104729N,2018EPJC...78..216H} and the anomalous magnetic 
moments of leptons \cite{2018JHEP...02..107D,2018JHEP...02..121A,
2017arXiv171201239G,2018arXiv180104729N} were also discussed. 
It needs to be noted that if there is indeed a massive subhalo in 
the local vicinity of the solar system, the local density of DM should 
be different from the canonical one of $\sim0.3$ GeV cm$^{-3}$ 
\cite{2016MNRAS.463.2623H}. For example, for a distance of 0.1 kpc and 
a subhalo mass of $10^6$ ($10^7$) M$_{\odot}$, the local density becomes 
$\sim4$ (19) times as large as the canonical value. Therefore, any 
direct detection limits based on the canonical local density should 
be lower by such a factor. While the model predictions may still be 
consistent with the current data, the future direct detection and 
collider experiments can definitely probe a large portion of the 
model parameters, as illustrated in Ref.~\cite{2017arXiv171110995F}.

Other than the DM annihilation/decay, some models with additional exotic
physical processes have also been proposed. In Ref. \cite{2016arXiv161108384J}
a threshold interaction with a hypothetical light particle $X$ was employed
to explain the knee and also the peak of the DAMPE data. In Ref.
\cite{2017arXiv171207868J} a gluon condensation effect during the
hadronic interaction of nuclei was proposed. The future observations of
multi-messengers, including $\gamma$-rays, neutrinos, anisotropy of CREs, 
may help test both the DM models and these exotic physical models
(see e.g., \cite{2017arXiv171110995F,2017arXiv171203210Z}).

\section{Summary}

Recently the DAMPE collaboration reported the measurement of the
high-energy CRE spectrum with very high precision. Here the DAMPE 
experiment, data analysis, and the physical interpretations of its
results are reviewed. The DAMPE CRE data, which is by far the most precise 
in the TeV band, improves significantly the constraints on the model 
parameters to explain CREs. Remarkably, the DM annihilation/decay
models to explain the positron excess and CRE spectra can be largely 
excluded with joint efforts of AMS-02, DAMPE, Fermi, and PLANCK.
One may either need to finely tune the DM models with more complicated
assumptions or resort to astrophysical solutions to the positron
anomaly. A tentative peak structure around $\sim1.4$ TeV can be seen
in the CRE spectrum. If it is confirmed by more data, this would be a 
very interesting and important discovery which is unexpected before.
To produce such a spectral feature, the sources of CREs are in general
required to be 1) nearby (with a distance $\lesssim 0.3$ kpc) and
2) spectrally hard (or even quasi-monochromatic). The cold,
ultra-relativistic $e^+e^-$ wind from pulsars without modification
of energies of such wind particles, or the DM annihilation/decay in
a nearby subhalo are suggested to account for the data. We emphasize
again that the most important things to crucially address this issue
are to accumulate more data with DAMPE, and to improve the analysis 
method to reduce the systematic uncertainties.

\section*{Acknowledgements}
We thank Dr. Yi-Zhong Fan for helpful discussion.
This work is supported by the National Key Research and Development Program 
of China (No. 2016YFA0400200), National Natural Science Foundation of China 
(Nos. 11722328, 11773075, U1738205), the 100 Talents Program of Chinese 
Academy of Sciences, and the Youth Innovation Promotion Association 
of Chinese Academy of Sciences (No. 2016288). 

{}

\end{document}